\documentclass{article}

\usepackage{arxiv}

\usepackage[utf8]{inputenc} 
\usepackage[T1]{fontenc}    
\usepackage{hyperref}       
\usepackage{url}            
\usepackage{booktabs}       
\usepackage{amsfonts}       
\usepackage{nicefrac}       
\usepackage{microtype}      
\usepackage{graphicx}
\usepackage{longtable}
\usepackage{multicol}
\usepackage{capt-of}
\usepackage{amsmath}
\usepackage[authoryear]{natbib}

\title{{Dust of comet 67P/Churyumov-Gerasimenko collected by Rosetta/MIDAS}: classification and extension to the nanometre scale}

\author{
  T.~Mannel\thanks{Space Research Institute of the Austrian Academy of Sciences, Schmiedlstrasse 6, 8042 Graz, Austria. \texttt{thurid.mannel@oeaw.ac.at}}
  \textsuperscript{\normalfont{,}}\thanks{University of Graz, Universitätsplatz 3, 8010 Graz, Austria.}\hspace{3pt},
  \hspace{1pt}
  M.S.~Bentley\thanks{European Space Astronomy Centre, Camino Bajo del Castillo, s/n., Urb. Villafranca del Castillo, 28692 Villanueva de la Cañada, Madrid, Spain.}\hspace{4pt},
  \hspace{1pt}
  P.D.~Boakes\textsuperscript{\normalfont{1}},
	\hspace{1pt}
	H.~Jeszenszky\textsuperscript{\normalfont{1}},
	\hspace{1pt}
	P.~Ehrenfreund\thanks{Leiden Observatory, Postbus 9513, 2300 RA Leiden, The Netherlands.}
	\hspace{1pt}\textsuperscript{\normalfont{,}}\thanks{Space Policy Institute, George Washington University, 20052 Washington DC, USA.}\hspace{3pt},
	\hspace{1pt}
    C.~Engrand\thanks{Centre de Sciences Nucléaires et de Sciences de la Matière (CSNSM) CNRS-IN2P3/Univ. Paris Sud, Université Paris-Saclay, Bât. 104, F-91405 Orsay Campus, France.}\hspace{4pt},
	\hspace{1pt}
	C.~Koeberl\thanks{Department of Lithospheric Research, University of Vienna, Althanstrasse 14, 1090 Vienna, Austria.}
	\hspace{1pt}\textsuperscript{\normalfont{,}}\thanks{Natural History Museum, Burgring 7, 1010 Vienna, Austria.}\hspace{3pt},\\
	\hspace{1pt}
	\textbf{A.C.~Levasseur-Regourd}\thanks{UPMC (Sorbonne Univ.), CNRS/INSU, LATMOS-IPSL, Paris, France.}\hspace{4pt},
	\hspace{1pt}
	\textbf{J.~Romstedt}\thanks{European Space Research and Technology Centre, Future Missions Office (SREF), Noordwijk, The Netherlands.}\hspace{8pt},
	\hspace{1pt}
	\textbf{R.~Schmied}\textsuperscript{1},
	\hspace{1pt}
	\textbf{K.~Torkar}\textsuperscript{1},
	\hspace{1pt}
	\textbf{I.~Weber}\thanks{Institut für Planetologie, Universität Münster, Wilhelm-Klemm-Strasse 10, 48149 Münster, Germany.}\hspace{8pt}.  	
}

\begin{document}
\maketitle

\begin{abstract}
\textit{Context.}
The properties of the smallest subunits of cometary dust contain information on their origin and clues to the formation of planetesimals and planets. Compared to IDPs or particles collected during the Stardust mission, dust collected in the coma of comet 67P/Churyumov-Gerasimenko during the Rosetta mission provides a resource of minimally altered material with known origin whose structural properties can be used to further the investigation of our early Solar System.
\newline
\textit{Aims.}
The cometary dust particle morphologies found at comet 67P on the micrometre scale are classified and their structural analysis extended to the nanometre scale.
\newline
\textit{Methods.}
A novel method is presented to achieve the highest spatial resolution of imaging possible with the MIDAS Atomic Force Microscope on-board Rosetta. 3D topographic images with resolutions of down to 8\,nm are analysed to determine the subunit sizes of particles on the nanometre scale.
\newline
\textit{Results.}
Three morphological classes can be determined, namely (i) fragile agglomerate particles of sizes larger than about 10\,$\mathrm{\mu m}$ comprised by micrometre-sized subunits that may be again aggregates and show a moderate packing density on the surface of the particles; (ii) a fragile agglomerate with a size about few tens of micrometres comprised by micrometre-sized subunits suggested to be again aggregates and arranged in a structure with a fractal dimension less than two; (iii) small, micrometre-sized particles comprised by subunits in the hundreds of nanometres size range that show surface features suggested to again represent subunits. Their differential size distributions follow a log-normal distribution with means about 100\,nm and standard deviations between 20 and 35\,nm.
\newline
\textit{Conclusions.}
The properties of the dust particles found by MIDAS represent an extension of the dust results of Rosetta to the micro- and nanometre scale. 
All micrometre-sized particles are hierarchical dust agglomerates of smaller subunits. The arrangement, appearance and size distribution of the smallest determined surface features are reminiscent of those found in CP IDPs and they represent the smallest directly detected subunits of comet 67P.
\end{abstract}


\setcounter{footnote}{0}

\begin{multicols}{2}

	\section{Introduction}
	The process that forms planets, asteroids and comets is usually estimated to have started with collisional aggregation of the smallest dust particles, themselves products of earlier stellar evolution or condensation processes in our early Solar System~\citep{tielens_dust_2005,li_dust_2003, weidenschilling_formation_1993}. Due to their importance in the agglomeration process, these particles have been the focus of many observational and laboratory studies~\citep{blum_growth_2008}.
	However, it was not previously possible to investigate the microscopic	properties of nearly unaltered, individual particles with a known provenance. Former investigations were based on (1) remote observations~\citep{levasseur_dust_2008, Hayward_imaging_2000}, (2) laboratory measurement of returned cometary material~\citep{brownlee_comet_2006, zolensky_mineralogy_2006} and (3) linking interplanetary dust particles (IDPs) collected in the Earth’s stratosphere~\citep{wozniakiewicz_sorting_2013, zolensky_wild2_2008, bradley_idps_2007} and ultracarbonaceous Antarctic Micrometeorites (UCAMMs)~\citep{yabuta_ucamms_2017, dartois_ucamms_2013, duprat_ucamms_2010} 
	to cometary material. 
	
	The Rosetta mission to comet 67P/Churyumov-Gerasimenko (hereafter 67P) provided the first opportunity to sample the dust and gas environment of the inner coma of a comet during a 2 years period around its perihelion passage in August 2015. Due to the low relative spacecraft-nucleus speeds, dust particles were collected by various instruments with only small degrees of alteration. One of the instruments on-board, MIDAS (Micro-Imaging Dust Analysis System,~\citep{bentley_lessons_2016, riedler_midas_2007}), carried the first Atomic Force Microscope (AFM) launched into space, and was specifically designed to probe the properties of the smallest dust particles at the micro- to nanometre scale. 
	
	Investigations of the dust particles collected by MIDAS are reported in~\citet{bentley_morphology_2016} and~\citet{mannel_fractal_2016}. The most important conclusion of these papers is the structural description of nearly unaltered cometary dust particles in the micrometre size range: all detected dust particles show hierarchical agglomerate character. 
	Here we introduce a classification of the particles analysed by MIDAS into three groups: (i) the large, about 10\,$\mathrm{\mu}$m sized particles with subunits packed in a moderately dense fashion that make up the majority of MIDAS' collection; (ii) one large but extremely porous particle; (iii) the small, a few micrometre-sized particles. 
	Whilst groups (i) and (ii) were already analysed in~\citet{bentley_morphology_2016} and~\citet{mannel_fractal_2016}, a characterisation of the small particles is presented in this paper. In particular, it will be shown that the small particles are not only comprised of some hundred nanometre-sized subunits as introduced in~\citet{bentley_morphology_2016}, they also show surface features 
	following a log-normal differential size distribution with means measuring about 100\,nm and standard deviations between 20 and 35\,nm.
	To obtain this resolution a novel imaging technique was developed that is described in Sect.~\ref{sec:methods}. 
	High resolution images of a well-preserved dust particle are shown in Sect.~\ref{sec:results_nm_scale} and the size distributions of the subunits and the smallest identifiable surface features are presented. In Sect.~\ref{sec:discussion_smallest_particles} they are discussed and compared to the results of IDP and UCAMM 
	analysis, findings of the Stardust mission, and to other Rosetta results.
	The classification of MIDAS results is shown in Sect.~\ref{sec:results_classification} and set into the frame of the results obtained about dust during the Rosetta mission in Sect.~\ref{sec:discussion_classification}. Sect.~\ref{sec:conclusions} gives a brief summary of the main findings.

	
	\section{Methods}\label{sec:methods}
	
	The MIDAS instrument on-board the Rosetta comet orbiter collected dust particles and imaged them with an AFM. A description of the instrument is presented by~\citet{riedler_midas_2007} and an overview of its operation and imaging modes is given in~\citet{bentley_lessons_2016}. Like every AFM, MIDAS used sharp tips to raster the dust surface in order to obtain high resolution 3D images of nearly unaltered cometary dust. 
	The finite width of the tips became a limiting factor when trying to access the smallest features of the dust particles~\citep{bentley_lessons_2016}. 
	Therefore a 'reverse imaging mode' was developed, in which an on-board tip calibration sample with sharp spikes  was used to probe cometary dust particles that had accumulated on the AFM tips during previous scans. 
	This tip calibration sample was used primarily to image the apex of the tip to determine its shape. It consisted of an array of spikes sharper than the tips with a half-opening angle of 25$^{\circ} \pm $ 5$^{\circ}$, 700\,nm height, a distance between neighbouring spikes of 2.12\,$\mathrm{\mu m}$ and of 3\,$\mathrm{\mu m}$ in the diagonal direction~\citep{cal_sample}. 
	
	As an AFM image is a convolution of the real structure and the tip shape, the tip width and shape can dominate the real resolution in certain situations, in particular if a steep feature smaller than the tip apex is imaged. Because the calibration spikes are sharper than the MIDAS tips, and there are many thousands of them available on the tip calibration target in case one gets blunt or contaminated, the ultimate resolution attainable with MIDAS is greatly improved by this reverse imaging mode.
	
	During MIDAS normal imaging mode (a dynamic intermittent contact mode~\citep{bentley_lessons_2016}) only low forces should be applied to the sample. Nevertheless, many of the cometary dust particles imaged fragmented, probably due to a high fragility, and dust was removed from the target or (partially) stuck to the tip. This pick-up of dust could either happen for whole particles at once, or via subsequent pick-up of smaller fragments. The adhering dust to the tip	was observed by the regular tip images acquired using the tip calibration sample. 
	Thus, the particles seen in the tip images have potentially undergone several modifications – on impact with the target, on pick-up by the tip, and in any subsequent scans	with this tip.
	A history of tip usage during and after pick-up is thus given in Sect.~\ref{history} in the appendix.
	In general, the overall particle surface structure might be altered and thus is not investigated in greater detail. It is instead assumed that the particles studied are aggregates of subunits which have a higher internal strength then the parent particle and thus their sizes and shapes might be still pristine~\citep{hornung_assessment_2016, skorov_dust_2012}.
	
	After successful image acquisition in the reverse imaging mode, a 3D image of the picked-up particle surface with resolutions of typically 15\,nm or 8\,nm is available.
	Following the methods used previously in~\citet{mannel_fractal_2016} and~\citet{bentley_morphology_2016}, the visible surface features of the particles were then identified by visual inspection of the topographic images and their 3D representation. Even the sharper tip calibration spikes cannot penetrate deeply between the individual features, and material apparently lying between the features was neglected as it cannot be fully imaged. These areas could either be further features hidden by an upper layer, an undefined matrix material, or features with a much smaller radius, which cannot be resolved due to the resolution limit. To prevent incorrect identification of subunits close to the resolution limit, no features with less than 9 pixel were marked.

	
	\section{Definition of language}
	
	The structural description of cometary dust is complex and benefits from a clearly defined vocabulary. A unified dust classification scheme of the Rosetta dust results with a well-defined vocabulary is presented in~\citet{guettler_synthesis_2019}. Here we use corresponding terminology, in particular:
	
	\begin{itemize}
		\item A \textbf{particle} is a subordinate term that can be applied either to any form of dust agglomerate, fragment, subunit, etc.
		
		\item A \textbf{grain} is the smallest building block of the dust, also referred to as \textbf{fundamental building block}. It is identified, e.g., based on its mineralogy or high material strength.
		
		\item An \textbf{agglomerate} consists of structurally distinct, smaller parts which can be, but do not have to be, different in properties. These smaller parts are termed \textit{subunits}. Each subunit can again be an agglomerate, but it can also be a grain. In the literature, the word \textit{aggregate} is often used synonymous to agglomerate, although there is a tendency to use agglomerate for loosely packed material and aggregate for more consolidated, stronger material (e.g.~\citet{nichols_agglomerate_2002}).
		
		\item A \textbf{fractal particle} is a hierarchical \textit{agglomerate} whose \textit{subunits} are arranged following a statistical order that can be described by a fractal dimension~\citep{mannel_fractal_2016, meakin_fractal_1991}. Although it is conceivable that cometary particles can have fractal dimensions Df~>~2, to date only dust particles with fractal structures of Df~<~2 have been detected with certainty~\citep{fulle_dust_2017, mannel_fractal_2016}.
		
		\item In the particular case of MIDAS, a \textbf{solid particle} is defined as one that does not show fragmentation or major surface features like deep trenches between bulbous units if scanned with sufficient resolution.
		
	\end{itemize}
	
	Furthermore, the \textbf{sizes} of particles and subunits are given as their equivalent diameters, i.e. the diameter of a disc with the same area as measured for the feature projected onto the x-y-plane.

	
	\section{Results}\label{sec:results}
	
	\subsection{Dust features on the nanometre scale}\label{sec:results_nm_scale}

\begin{figure*}
		\centering
		\includegraphics[width=17cm]{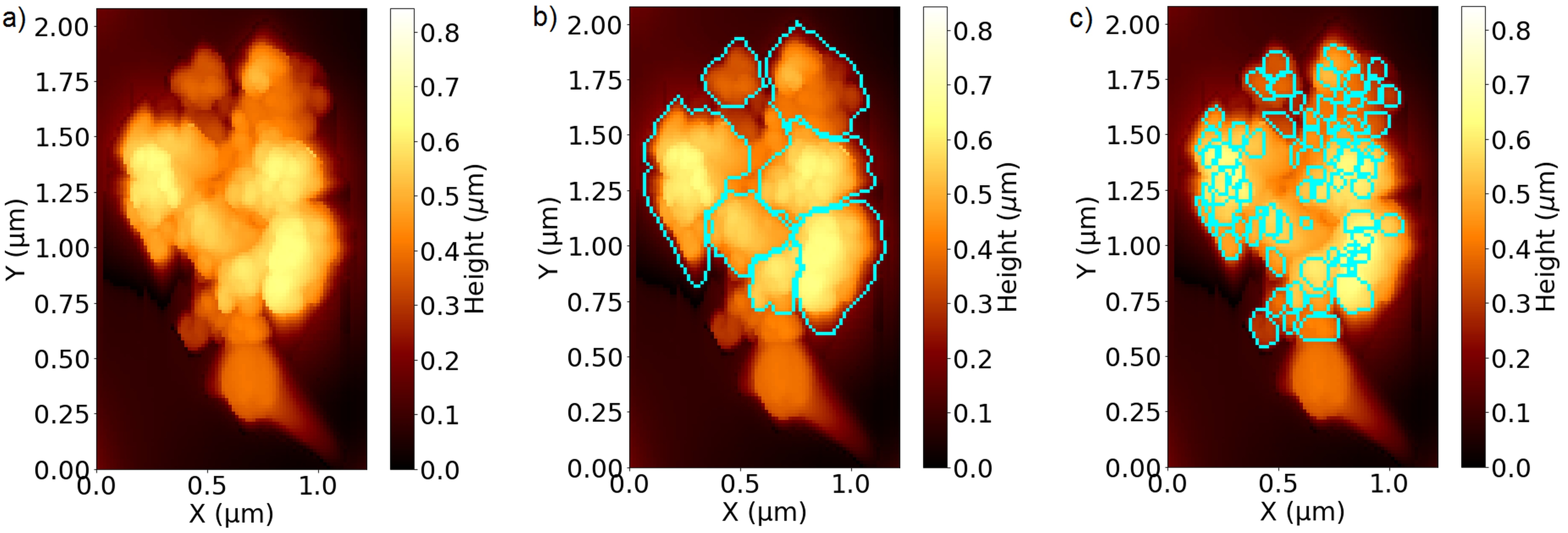}
		\caption{Image of Particle G, a 1\,$\mathrm{\mu m}$ sized particle scanned with MIDAS' reverse imaging mode on the 08 December 2015
			with a resolution of 15\,nm per pixel. The smooth, round shape at the bottom with a fading line to the bottom right corner is the tip with which the particle was picked-up, above sits the particle with well visible sub-structures. 
			(a) shows the particle itself, (b) the larger subunits, (c) the smallest identifiable features.}
		\label{fig_small_particle_15}
	\end{figure*}

\begin{figure*}
			\centering
			\includegraphics[width = 12cm]{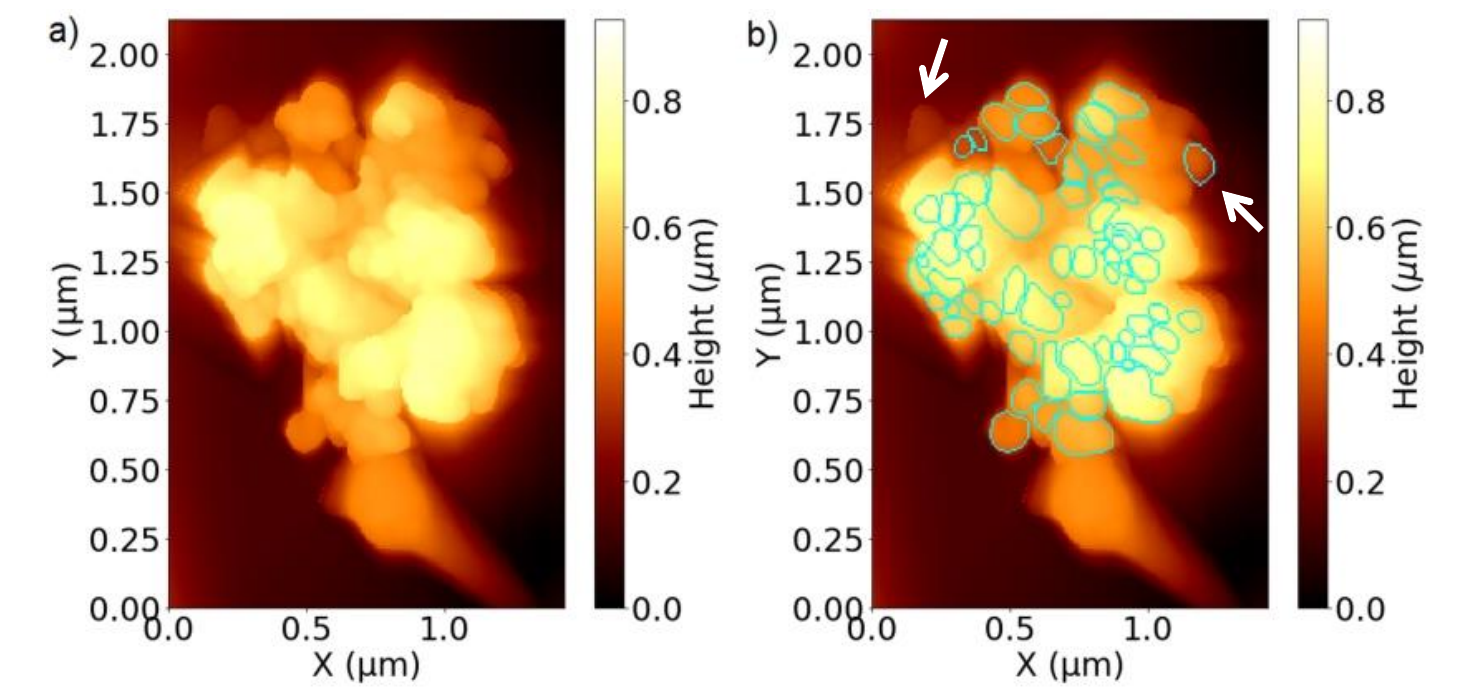}
			\caption{Image of the same 1\,$\mathrm{\mu m}$ sized particle G as shown in Fig.~\ref{fig_small_particle_15}, but this time scanned with a resolution of 8\,nm on 11 May 2016. 
				(a) shows the particle itself, (b) the smallest identifiable features. Compared to Fig.~\ref{fig_small_particle_15}, the colour scale has to be fully exploited as some lower-lying subunits were imaged (indicated by the white arrows).}
			\label{fig_small_particle_8}
		\end{figure*}

	One particle, called particle G, was collected during perihelion and was later picked-up by a tip (for a more precise description of its collection time and scan history see Sect.~\ref{history} in the appendix). It is expected to be rather unaltered due to its consistent structure in repeated scans (e.g., the images shown in Figs.~\ref{fig_small_particle_15} and~\ref{fig_small_particle_8}). 
	It was scanned seven times with resolutions of 15\,nm and 8\,nm between its first detection on 12 November 2015 and its last scan on 25 May 2016.	
	The scan that allows the best identification of the particle structure was taken on 08 December 2015. It has 15\,nm resolution and a crop of the interesting region is shown in Fig.~\ref{fig_small_particle_15}.
	The highest resolution scan with the least artefacts had a resolution of 8\,nm, was taken on 11 May 2016 and a crop of the particle is shown in Fig.~\ref{fig_small_particle_8}.
	The full scans together with key metadata can be found in~{Sect.~\ref{sec:raw}}.


	Fig.~\ref{fig_small_particle_15} (a) shows the 15-nm-scan taken on 08 December 2015. The particle is the open, flocculent structure  in the centre and measures about 1$\,\mathrm{\mu m}$ in diameter. It is adhering to the tip visible in the bottom right part of the image. The smooth, round shape is the tip apex, the straight line diminishing in height to the bottom right corner is a structure supporting the tip. 
	It is directly visible that particle G is an agglomerate of several large features that again show distinct surface features. The larger features are clearly separated and are thus treated as subunits comprising the particle. Whether or not their smaller surface features are also related to subunits is complex to decide based on MIDAS topographic data alone and will be discussed in Sect.~\ref{sec:discussion_smallest_particles}.
	
	Fig.~\ref{fig_small_particle_15} (b) shows the same image as Fig.~\ref{fig_small_particle_15} (a) but with the subunits outlined in cyan. As described in Sect.~\ref{sec:uncertainties} of the appendix the outer rim of the particle may show artificial broadening due to tip-sample-convolution depending on the steepness of the particle. 
	The large subunit on the left shows the most severe case: it is elongated to the bottom side due to said tip-sample-convolution. 
	
	Fig.~\ref{fig_small_particle_15} (c) shows the same image as the previous panels, but with the surface features marked in cyan. The subunits marked in (b) seem to be covered by the surface features or, if the surface features are subunits, the larger units may be completely built of them. However, it is not possible to mark all those surface features, e.g. due to limited discriminatory power related to the resolution, due to them seemingly covering each other, or being too close at the border. Thus, the marked features in Fig.~\ref{fig_small_particle_15} (c), as well as those marked in~Fig.~\ref{fig_small_particle_8} (b), are just a selection of the best visible features. Their spatial density and arrangement cannot be completely mapped.
	
	Fig.~\ref{fig_small_particle_8} (a) shows again particle G, the same particle as presented in Fig.~\ref{fig_small_particle_15}, however, this time scanned with 8\,nm resolution and 5 months later, on 11 May 2016. The structure of particle G remained unchanged, meaning that there was no substantial alteration during the scans, e.g. compression, fragmentation or displacements of subunits. 
	Additionally, the particle shows no change despite the 5 months long storage in the instrument at a temperature well above 20$^\circ$C, possibly up to 35$^\circ$C\footnote{The lowest temperatures measured in the instrument during those five months were rarely below 20$^\circ$C, while the temperature sensor closest to the dust particle measured most of the time temperatures around 35$^\circ$C.}.
	This suggests that the particle did not contain volatile materials when detected by MIDAS for the first time. It is however possible that the particle contained volatile materials that evaporated upon ejection from the nucleus, the travel through the coma, or in the three-month period between the collection and the detection by MIDAS. 
	
	The preservation of the structure indicates that particle G stayed relatively unaltered at least after pick-up by the tip. 
	Comparing Figs.~\ref{fig_small_particle_15} and~\ref{fig_small_particle_8}, the particle in Fig.~\ref{fig_small_particle_8} is slightly stretched horizontally and compressed vertically relative to the particle in Fig.~\ref{fig_small_particle_15}. The effect on the measured size of the particle is negligible as the measured equivalent diameter of particle G is $1213^{+32}_{-390}$\,nm and $1255^{+37}_{-460}$\,nm in the 15 and 8\,nm resolution scans, respectively, and is thus similar in the range of its uncertainties. The effect on the sizes of the smaller subunits is even smaller, thus the inaccuracy due to the stretch/compression can be neglected. 
	The slight change in shape of the particle in the different scans is an effect due to the longer scanning time of the higher resolved scan (7 hours versus 22 hours), making it more prone to piezo drift due to temperature changes. This is especially strong in the slower scanning direction (the horizontal direction in Figs.~\ref{fig_small_particle_15} and~\ref{fig_small_particle_8}). More details about this artefact are given in appendix Sect.~\ref{sec:raw}). 
	
	The scan shown in Fig.~\ref{fig_small_particle_8} (b) reveals additional, deeper lying features at the rim of the particle that are indicated by white arrows. The colour scale needed to be stretched and thus especially more shallow features cannot be easily recognized in the printed figure. Also, the rims of the particle in Fig.~\ref{fig_small_particle_8} show a stronger broadening which might be an effect caused by the use of different spikes on the calibration target that might have had distinct shapes and aspect ratios. 
	
	Fig.~\ref{fig_small_particle_8} (b) presents the same image as Fig.~\ref{fig_small_particle_8} (a) but this time with the surface features marked in cyan. In comparison to the features outlined in Fig.~\ref{fig_small_particle_15} (c), the majority of features was found in both scans. However, some features were only found in the 15\,nm resolution scan due to its slightly lower broadening of the particle rims, and some features were only found in the 8\,nm scan due to its better resolution.
	
	The cumulative size distributions of all marked subunits and features are given in~Fig.~\ref{fig:cum_sd_small_15}, \ref{fig:cum_sd_small_8} and~\ref{fig:cum_sd_large_15}. All sizes are listed in Tab.~\ref{table:meta_8} and their uncertainties were determined as described in Sect.~\ref{sec:uncertainties}. 
		The related differential size distributions can be expected to follow a log-normal function~\citep{wozniakiewicz_sorting_2013, rietmeijer_size_1993}.
		For the given measurement data the trend of the size distribution is easier to determine in the cumulative representation as the related uncertainties can be handled in a more convenient way~(see Sect.~\ref{sec:uncertainties}).
		Thus, the cumulative subunit sizes are fitted with the integrand of a log-normal distribution
		with the related mean values and standard deviations denoted as as $\mu_{log}$ and $\sigma_{log}$.
		All performed fits passed a Kolmogorov-Smirnov test (KS test). For details about the fitting routine and the KS test, see Sect.~\ref{sec_app_fit}.

		Fig.~\ref{fig:cum_sd_small_15} shows the sizes of the small surface features on particle G scanned with 15\,nm resolution (see Fig.~\ref{fig_small_particle_15} (c)) fitted by the integrand of a log-normal distribution. 
		The arithmetic mean value of the data measures $100.25^{+1.01}_{-6.34}$\,nm, in good agreement with the fitted mean value of $\mu_{log} = 101.80 \pm 0.50$\,nm. The geometric mean value is with $97.04^{+0.99}_{-7.29}$\,nm slightly smaller, an expected behaviour as this metric puts less weight on the larger sizes. 
		This can be favourable if it is expected that the large subunit sizes lie on the trailing end of a log-normal distribution and should be weighted less.
		The standard deviation is found to be $\sigma_{log} = 23.97 \pm 0.64$\,nm, and the minimal and maximal subunit sizes are $52^{+6}_{-26}$\,nm and $183^{+14}_{-139}$\,nm.

	\begingroup
		\centering
		\includegraphics[width=\linewidth]{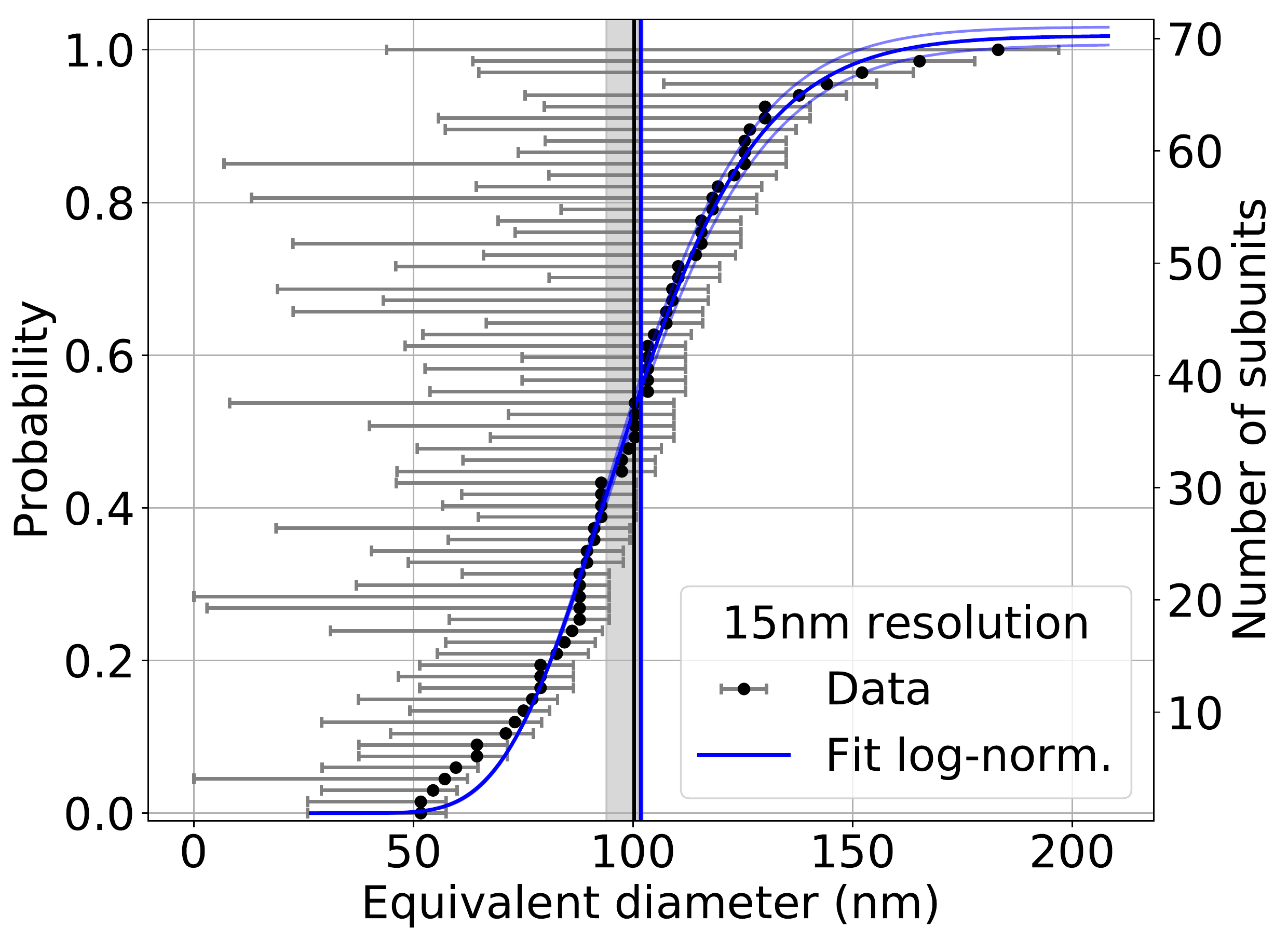}
		\captionof{figure}{Cumulative size distribution of the subunits of particle G identified in the 15\,nm resolution scan (Fig.~\ref{fig_small_particle_15}~(c)). On the left the probability that a subunit is smaller than the indicated value is shown, on the right the number of detected subunits. 
				The log-normal fit is shown in blue together with its uncertainty interval in light blue. The vertical lines denote the arithmetic (black) and fitted (blue) mean values with shaded uncertainty intervals.}
			\label{fig:cum_sd_small_15}
	\endgroup
		
	\begingroup
		\centering
		\includegraphics[width=\linewidth]{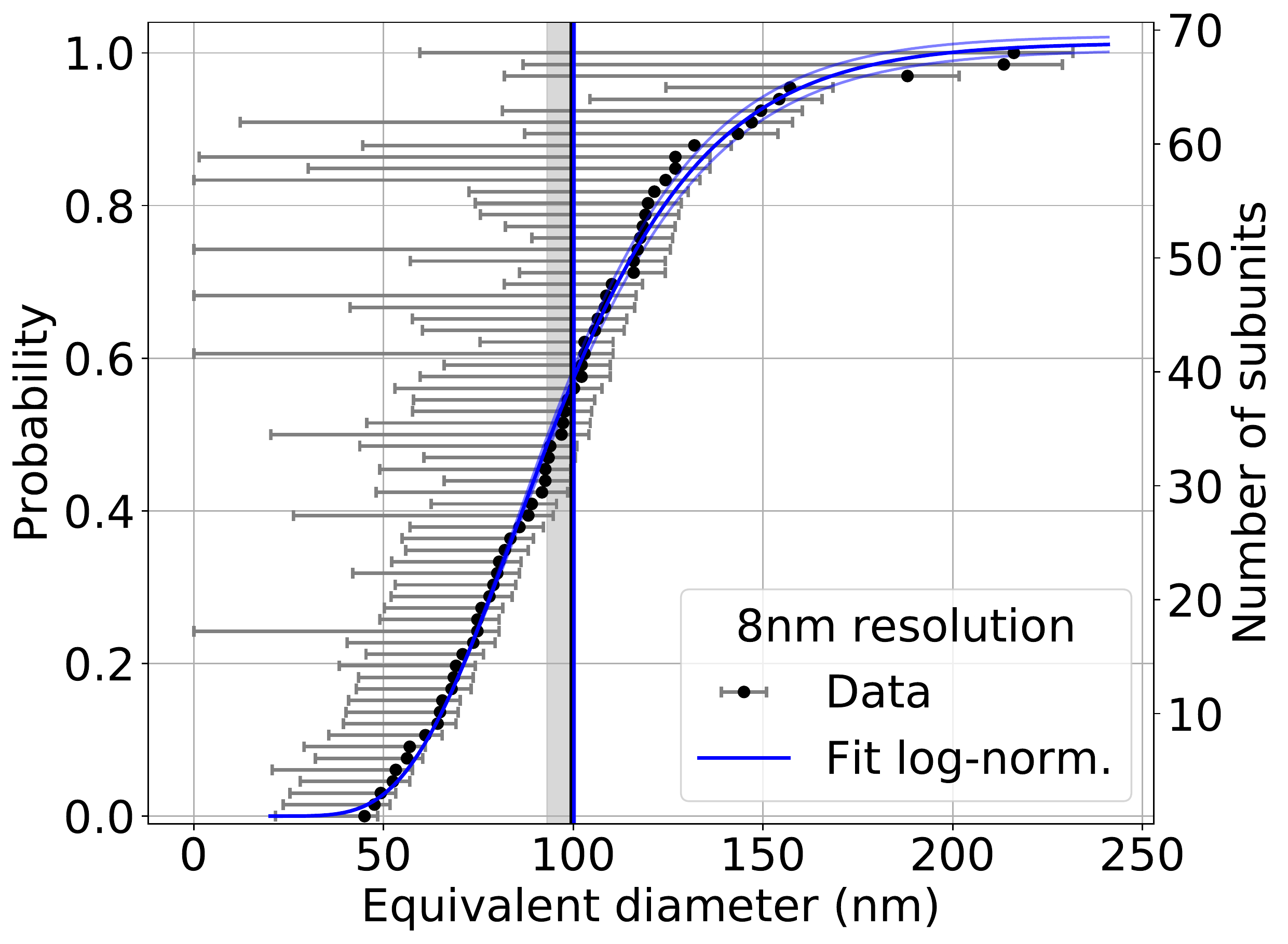}
		\captionof{figure}{Cumulative size distribution of the subunits of particle G identified in the 8\,nm resolution scan (Fig.~\ref{fig_small_particle_8}~(b)). On the left the probability that a subunit is smaller than the indicated value is shown, on the right the number of detected subunits. 
			The log-normal fit is shown in blue together with its uncertainty interval in light blue. The vertical lines denote the arithmetic (black) and fitted (blue) mean values with shaded uncertainty intervals.}
		\label{fig:cum_sd_small_8}
	\endgroup

				Fig.~\ref{fig:cum_sd_small_8} shows the small feature sizes on particle~G scanned with 8\,nm resolution (see Fig.~\ref{fig_small_particle_8} (b)) together with their fit.  
				The arithmetic mean value of $99.49^{+0.89}_{-6.41}$\,nm is in agreement with the fitted mean value of $\mu_{log} = 100.12 \pm 0.57$\,nm. The geometric mean value measures $93.79^{+0.85}_{-6.86}$\,nm.
				The influence of the two largest subunits sizes on the mean values is clearly reflected in the geometric mean being much smaller than the arithmetic mean.
				Comparing the mean values determined for the 15\,nm and 8\,nm scans the arithmetic mean values are in agreement, the fitted mean values are similar. 		
				The standard deviation of the 8\,nm scan is with $\sigma_{log} = 34.51 \pm 0.76$\,nm a bit broader than the standard deviation found for the 15\,nm resolution scan. This indicates a broader distribution, probably caused by the measured subunit sizes spanning a larger size interval between $45^{+3}_{-23}$\,nm and $216^{+16}_{-157}$\,nm.

	\begingroup
		\centering
		\includegraphics[width=\linewidth]{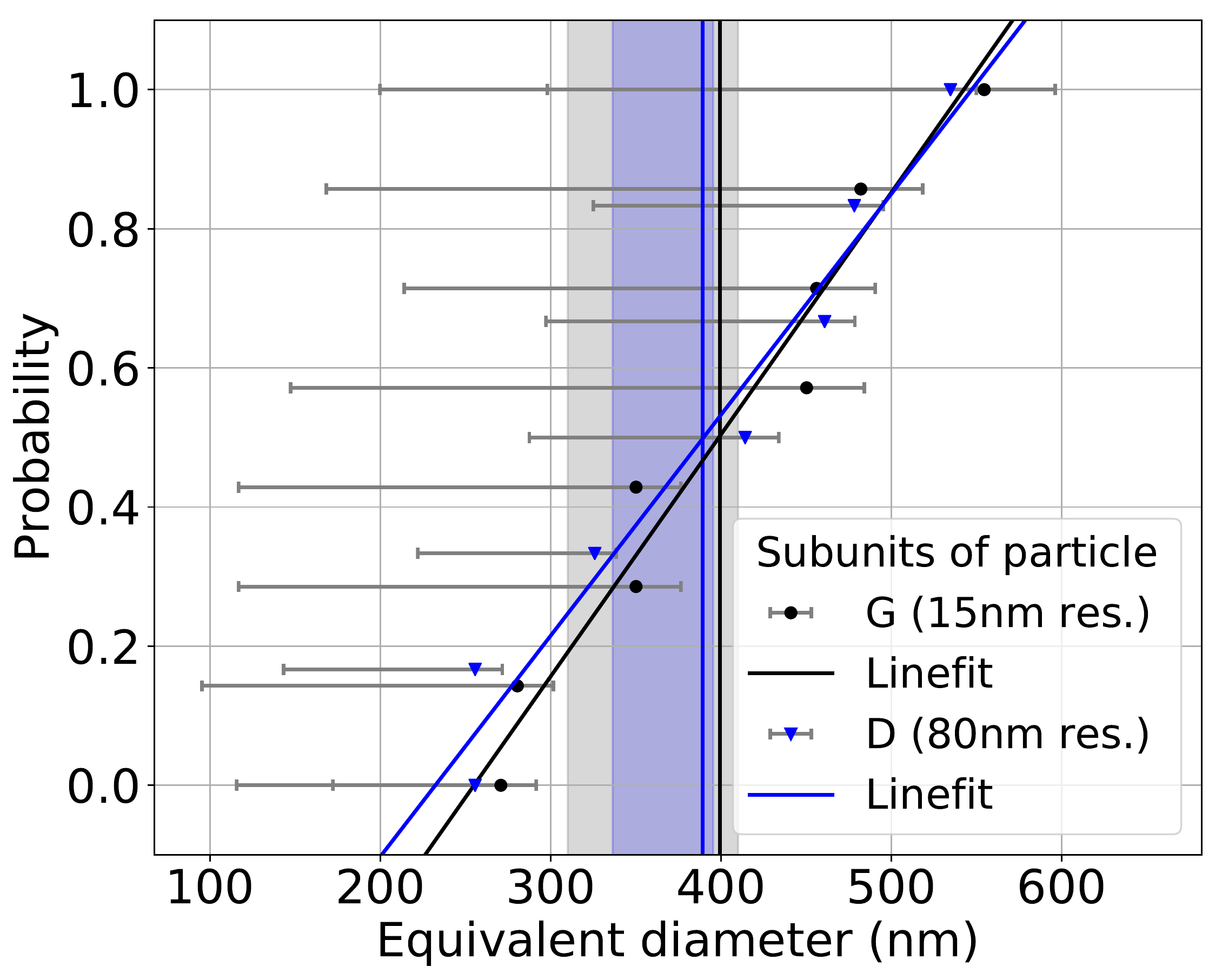}
		\captionof{figure}{Cumulative size distribution of the larger subunits of particle~G (Fig.~\ref{fig_small_particle_15}~(b), black dots) and D (\citep{bentley_morphology_2016}, blue triangles). The line fits of the data are shown for particle~G in black and particle~D in blue, the arithmetic means of the measured sizes are given as vertical lines with shaded uncertainty intervals.}
		\label{fig:cum_sd_large_15}
	\endgroup
			
					The distribution of the larger subunits of particle G as marked in Fig.~\ref{fig_small_particle_15} (b) is shown in black in Fig.~\ref{fig:cum_sd_large_15}. The sizes lie between $271^{+21}_{-155}$\,nm and $555^{+42}_{-355}$\,nm, have an arithmetic mean value of $399^{+11}_{-89}$\,nm and a geometric mean value of $388^{+ 10}_{-87}$\,nm. 
					Due to the low statistics, a discrimination between a log-normal distribution or a simple line is not possible. The KS test was passed best by the fit of a line with a slope of $(3.5 \pm 0.3)\cdot10^{-3}$\,nm$^{-1}$.

					The	sizes are in very good agreement with the sizes found for the subunits of particle D, an about 1\,$\mathrm{\mu}$m sized particle collected pre-perihelion~\citep{bentley_morphology_2016}.
					They are shown in blue in Fig.~\ref{fig:cum_sd_large_15}, lie between $256^{+50}_{-118}$\,nm and $535^{+24}_{-245}$\,nm, show an arithmetic mean value of $389^{+6}_{-53}$\,nm and a geometric mean value of $375^{+ 7}_{-52}$\,nm. The fitted line has a slope of $(3.2 \pm 0.3)\cdot10^{-3}$.
				In contrast to particle G, particle D was only imaged with a resolution of 80\,nm and thus no surface features on these subunits could be seen.
					The detection of similar-sized subunits at different resolutions indicates that the determined subunit size is independent of the image resolution. Finding subunits of similar sizes in distinct particles also underlines the suggested hierarchical dust structure with characteristic subunit size regimes~\citep{levasseur_dust_2018, bentley_morphology_2016}.
				As all high resolution scans of MIDAS show the same kind of surface features as particle G, and the subunit sizes of particle~D and particle~G match very well, it is possible that also particle D, as well as all subunits of the larger particles~E and~F visible in Fig.~\ref{fig:3ds}, also have surface features in the hundred nanometres size range. It can be hypothesized that many of the micrometre-sized particles of comet 67P have the approximately 100\,nm sized surface features visible on particle~G.
				
					In summary, the differential distributions of the small features of particle~G follow a log-normal distribution with a mean size about 100\,nm and a standard deviation between 20 and 35\,nm, where the detected subunits span a total size range between about 50 to 250\,nm.		
					The lower limit of feature sizes determined in particle G seems not to be resolution limited as the smallest feature size and mean values stay similar in differently resolved scans. The upper size limit is given by the transition to larger features (marked in Fig~\ref{fig_small_particle_15} (b)). Deep trenches separate the larger features such that they are clearly distinct and are suggested to be subunits. They exhibit the smaller features on their surfaces (marked in Fig.~\ref{fig_small_particle_15} (c)) which indicates that the larger subunits may be comprised by smaller ones.

				\subsection{Dust classification}\label{sec:results_classification}
				
				In an attempt to classify the MIDAS dust collection, three distinct particle classes can be identified.	
				
				\paragraph{(i) Large particles with moderate packing of subunits at the surface} 
				
\begin{figure*}
					\centering
					\includegraphics[width = 17cm]{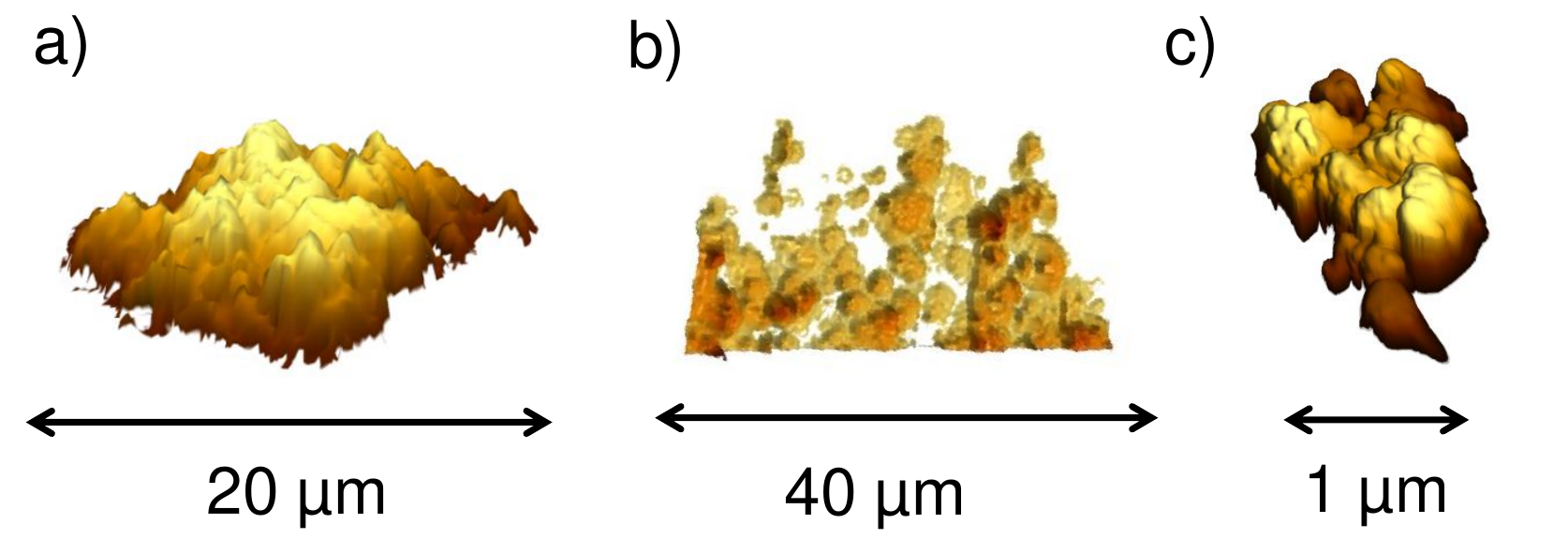}
					\caption{3D rendered images of MIDAS dust particles. (a) shows particle~F as example for a large agglomerate particle with its subunits packed in a moderately dense fashion on the surface~\citep{mannel_fractal_2016}. The source scan was taken on 14 October 2015 at 08:08:23~UTC with a resolution of 192\,nm. (b) shows the large porous agglomerate particle E. The data are taken from a scan on 18 January 2015 at 20:59:28~UTC with a resolution of 210\,nm. (c) shows particle G representative for the small particle class. The source scan was taken at 11 May 2016 at 12:08:03~UTC with a resolution of 8\,nm and is shown in Figs.~\ref{fig_small_particle_8} and~\ref{fig:raw_8}.}
					\label{fig:3ds}
				\end{figure*}

				The large (>~10\,$\mathrm{\mu m}$ in size) particles are agglomerates with subunit sizes around 1.5\,$\mathrm{\mu m}$~\citep{mannel_fractal_2016} that could potentially again consist of smaller subunits which were not seen due to the resolution limit in these scans. The subunits visible on the surface are neither packed in the densest possible fashion, but also not extremely loose; there are trenches visible that clearly separate the subunits, but those trenches are generally smaller than the typical size of the subunits, and in particular in comparison to the large porous particle one cannot see the target surface through the particle. 
				In contrast to smaller particles detected with MIDAS, the large particles with moderate packing fragment easily into their micrometre-sized subunits. This is interpreted as an indication for the strength keeping the subunits of the large particles together being lower than the strength binding the subunits of the small particles as it is expected that the effective tensile strength grows larger for smaller aggregates~\citep{hornung_assessment_2016, skorov_dust_2012}. 
				There was no case where a disintegrated particle showed fragments that differed from those observed on the surface of the particle and thus it is suggested that all large agglomerates internally consist of the same typical subunits as visible at the surface. There are no observations indicating that the large particles might have a solid core that is just coated with the observed subunits, or that there are euhedral crystalline parts, meaning crystalline material with clear cut, recognisable faces. 
				The majority of MIDAS detections are large compact agglomerates. An example is particle~F, analysed in~\citet{mannel_fractal_2016}, and shown as rendered 3D image in Fig.~\ref{fig:3ds} (a).

				\paragraph{(ii) The large porous particle}
				
				The large (>~10\,$\mathrm{\mu m}$ in size) porous particle, called particle E, consists of similar sized and shaped subunits as visible on the surface of the large particles with moderate packing~\citep{mannel_fractal_2016}. However, as visible in the rendered 3D image of the particle presented in Fig.~\ref{fig:3ds}~(b), they are extremely loosely assembled. 
				As derived in~\citet{mannel_fractal_2016} the structure shows a degree of ordering that can be described with a fractal dimension less than two. It is assumed that this structure is characteristic for dust particles built in a Cluster-Cluster-Agglomeration growth process in the solar nebula~\citep{blum_growth_2008, dominik_dust_2007}. As shown in~\citet{fulle_fractal_2017} and~\citet{mannel_fractal_2016} the porous large particle might have been preserved since the early Solar System, being the structurally most pristine dust particle ever imaged. 
				It is suggested that all dust particles found in comets started with this fractal structure, but the majority was subsequently densified and is now forming the group of particles with a moderate packing of subunits~\citep{blum_evidence_2017, fulle_fractal_2017}. MIDAS detected one large porous particle, particle~E, however, the GIADA instrument (Grain Impact Analyser and Dust Accumulator~\citep{colangeli_GIADA_2007}) demonstrated that there is a whole population of these particles~\citep{fulle_dust_2017, fulle_density_2015}.
				
				\paragraph{(iii) Small particles}\label{sec:small_particles}
				
				The small (about 1 to a few micrometres in size) particles show bulbous shaped subunits with a surface packing density suggested to be comparable to that of the large particles with moderate packing fashion of the subunits. The 1\,$\mathrm{\mu m}$ particles typically consist of subunits with several hundreds of nanometres in size. 
				As the individual small particles are similar in size and shape to the fragments of the large particles, it is suggested that they might fall into the same population. One possibility could be that the individually collected small particles are in fact fragments from larger particles that separated before collection, e.g. during ejection from the nucleus or their travel through the cometary coma. Another option could be that they were not integrated into the larger particles during formation in the early Solar System, but nevertheless were incorporated into the material of the cometary nucleus.	
				In contrast to the large particles with moderate packing of subunits at the surface, no small particle showed fragmentation or strong alteration during scanning. This is indicative of a higher internal strength keeping the smaller particles together.
				In general, individual small particles are scarce in MIDAS detections, and no individual particle of less than 1\,$\mathrm{\mu}$m was detected. Possible reasons are discussed in Sect.~\ref{sec:discussion_lack_solid}.
				
				Particle G shown as rendered 3D version in Figure~\ref{fig:3ds} (c), as well as particle D presented in~\citet{bentley_morphology_2016}, are exemplary particles for the small particle class. All small particles and fragments of larger particles scanned with a sufficiently high resolution show subunits less than hundred nanometres in size. Thus, it is suggested that the 100\,nm sized surface features are common to all cometary dust of comet 67P, e.g. also the subunits of particle~E in Fig.~\ref{fig:3ds} (b) or particle~F in Fig.~\ref{fig:3ds} (a) might well show these features if scanned with a higher resolution. The same is suggested for the small particle~D presented in ~\citet{bentley_morphology_2016}.

				
				\section{Discussion}\label{sec:discussion}	
				
				\subsection{Comparison of the smallest subunit sizes found in comet 67P and other samples}\label{sec:discussion_smallest_particles}
				
				The investigation of the dust of comet 67P in the reverse imaging mode of MIDAS was very successful. It was possible to determine the surface structure of the smallest individually collected particles down to the less than 100\,nm scale. The approximately 1\,$\mathrm{\mu m}$ sized particles show clearly separated features that are interpreted as subunits with mean sizes close to 400\,nm.
				Most interestingly, these subunits again show features with sizes following a log-normal distribution with a mean about 100\,nm in size. 
				With the purely topographic information provided by MIDAS, it cannot be conclusively decided if these smallest features are subunits or just surface features. 
				For such a determination higher resolution topographical scans, as well as information on the material properties, such as compositional heterogeneity, would be needed. However, a suggestion can be made based on a comparison to other Rosetta measurements and to typical sizes found for subunits identified in other cometary material.
				
				The best sources for subunit sizes of cometary dust are the returned Stardust samples~\citep{brownlee_comet_2006} from comet 81P/Wild~2, investigations of Chondritic Porous Interplanetary Dust Particles (CP IDPs) collected in the stratosphere~\citep{flynn_organic_2013} and Antarctica~\citep{noguchi_dust_2015}, as well as UCAMMs~\citep{duprat_ucamms_2010} gathered on the Antarctic continent.
				All those materials are susceptible to alteration between their time of release from the comet and their investigation in the laboratory. Stardust samples had to survive a high-velocity capture, IDPs and UCAMMs 
				had a long Solar System sojourn and were potentially altered during their passage of Earth's atmosphere. 
				Keeping those shortcomings in mind, those particles nevertheless represent a great resource to study their parent body properties such as, e.g., their subunit sizes. 
				
				\paragraph{CP IDPs and UCAMMs}
				Both materials 
				consist of sub-micrometre-sized subunits consisting of anhydrous materials, mainly olivine, pyroxene, a substance described as Glass with Embedded Metal and Sulphides (hereafter GEMS), and iron sulphide~\citep{flynn_organic_2013, dobrica_ucamms_2012}. 	
				A study by~\citet{wozniakiewicz_sorting_2013} investigated the size distribution of the subunits of CP IDPs. The dust agglomerates were disaggregated and the sizes of the resulting fragments measured. Thus, the sizes refer not primarily to compositionally distinct regions (that cannot be determined by MIDAS either), but to the sizes of the consolidated subunits (a quantity accessible to MIDAS).
					Over 5600 subunits of four CP IDPs were analysed and their size distributions found to follow log-normal distributions~\citep{wozniakiewicz_sorting_2013}. The mean and standard deviation values of the fitted distributions were not determined, however, the cumulative size distributions were investigated by a graphical procedure that determined the geometric means to range between $68^{+6}_{-4}$\,nm and $306^{+10}_{-6}$\,nm and the standard deviations between about $40$\,nm and $200$\,nm~\citep{wozniakiewicz_sorting_2013}.
					The geometric mean value of the small features of particle G are, although on the smaller side, in agreement with the geometric mean values found for the CP IDPs.
					The distribution of the small features is narrower than that of the CP IDPs which could be an effect due to the subunit sizes of the CP IDPs covering a larger size range (from about tens of nanometres up to 1\,$\mu$m).		
					The narrow nature of the subunit size distribution in dust of comet 67P was already mentioned in earlier studies at larger size scales: on the one hand for the subunits of a few micrometres size comprising the 10\,$\mu$m-sized particles as detected by MIDAS~\citep{mannel_fractal_2016}; on the other hand for the tens of micrometre-sized subunits comprising the 100\,$\mu$m-sized clusters detected in COSIMA optical microscope images with a pixel resolution of 14\,$\mu$m~(Cometary Secondary Ion Mass Analyser~\citep{kissel_COSIMA_2007})~\citep{hornung_assessment_2016}. 
					The repeated detection of rather distinct subunit sizes supports the image of a hierarchical structure of cometary dust (see also Sect.~\ref{sec:hierarchy}).

					\citet{rietmeijer_size_1993} derived the subunit size distribution for one CP IDP. The particle was imaged with transmission and scanning electron microscopy and its subunits visually identified. The particle showed about 100 granular units, subunits of carbon-rich chondritic to carbonaceous composition and polyphase units, most frequently about 100\,nm in size~\citep{rietmeijer_size_1993}. 
					Their diameters range from 64\,nm to 7580\,nm with a mean of 585\,nm, all sizes given with a relative uncertainty of 10 percent.
					Their size distribution follows a log-normal distribution typical for a size-sorting process~\citep{rietmeijer_size_1993}. It is polymodal with overlapping normal distributions described by means between 128\,nm and 3360\,nm. 
					The size range covered by the subunits of the investigated CP IDP is much larger than found in particle G, in particular because it trails to much larger sizes. Only the normal distribution with the smallest mean of 128\,nm comes close to the observed mean value about 100\,nm of the smallest features of particle G, but since no standard deviation for this distribution is given it is uncertain how well the size distribution of particle G and the investigated CP IDP would match.

					Additionally, \citet{rietmeijer_size_1993} detected about 400 nanocrystals among the constituents of the granular units of two CP IDPs. They had sizes between 1.4 and 636\,nm that follow log-normal and log-log-normal distributions with means between 3.1\,nm and 49.6\,nm and standard deviations between 0.5 and 7.2\,nm.
					It is conceivable that also the herein investigated subunits of particle G contain similar nanocrystals or particles as small as some nanometres that were not resolved in the 8\,nm resolution scans of MIDAS. If such nanocrystals should be treated as smallest subunits or if they are constituents fused together to form the smallest subunits would need a separate discussion if they were existent. 
								
					\citet{dobrica_ucamms_2012} analysed three UCAMMs and investigated the apparent sizes of grains visible in 80\,nm thin sections. They measured 550 mineral subunits (olivines, pyroxenes and sulfides) and found sizes ranging from 15\,nm to 1.1\,$\mathrm{\mu m}$ with a geometric mean of about 138\,nm and an uncertainty of the size measurements of 5 percent. As no fits are available to date, it is not possible to determine how well the size distributions would fit in the range measured by MIDAS. However, as the size range encloses that of the features found in MIDAS particles, the possibility that the size distributions are in agreement is given and could be tested in future projects.

					In summary, judging by the size of the detected features in CP IDPs, the smallest subunits found in MIDAS dust particles could correspond to the smallest subunit sizes derived for CP IDPs. 
					However, the herein determined size distributions are narrower, in particular they trail less to larger values. This could be an effect due to low statistics or a true difference between the samples.

				\paragraph{Stardust measurements}
				The majority of the particles collected by Stardust are olivine and pyroxene silicates with solar isotopic compositions, which suggests an origin in our Solar System rather than an interstellar provenance. These polymineralic particles dominate over those made of a single mineral even down to sizes
				less than 100 nm, indicating that the dust composition is surprisingly consistent at
				different scales and that the smallest subunits of the dust may be as small as
				tens of nanometres~\citep{hoerz_impact_2006, zolensky_mineralogy_2006}.
					The sizes of these smallest single mineral impactors are similar to those of the nanocrystals determined by~\citep{rietmeijer_size_1993}. As discussed above, these might also be existing in MIDAS dust particles and they could be fused into the 100\,nm sized features.
				\citet{price_size_2010} and~\citet{wozniakiewicz_grain_2012} investigated the sizes of the small, less than 10\,$\mathrm{\mu m}$ sized particles that impacted the aluminum foils of the Stardust probe.
					The distribution peaks at about 175\,nm, however, if assuming that the particles are agglomerates of smaller subunits as indicated by their common polymineralic nature, then the subunit size distribution would peak at sizes below 100\,nm~\citep{price_size_2010}. 
					A study of over 450 particles that do not seem to be agglomerates, i.e. those that show single mineral impactors of silicate or sulfide, found geometric mean sizes of  $532^{+741}_{-310}$\,nm for the silicate particles and $406^{+491}_{-222}$\,nm for the sulfides~\citep{wozniakiewicz_sorting_2013}. Those sizes are notably larger than the 175\,nm (or less) found for the whole dataset. This large spread of subunit sizes could indicate a size distribution with a large width. 
					There are no fits of these size distributions available, however, the figures in~\citet{wozniakiewicz_grain_2012} and~\citet{price_size_2010} open the possibility that the differential sizes follow a log-normal distribution.
					With the smallest subunit sizes possibly between tens and hundreds of nanometres, the subunit size range found for MIDAS smallest features would be encompassed. The determination of the size distributions for the small Stardust particles and a detailed comparison to the distributions obtained for comet 67P could be the work of an interesting future project.

				\paragraph{Other Rosetta measurements}
				Although there was no instrument other than MIDAS on-board Rosetta that directly measured (sub-)micrometre sized dust particles, there were several instruments that indirectly detected the presence of smallest dust. The remote sensing instruments Alice (ultraviolet imaging spectrometer~\citep{stern_alice_2007}) and VIRTIS  (Visible and InfraRed Thermal Imaging Spectrometer~\citep{coradini_virtis_2007}) suggest that the dust properties change during some dusty outbursts. \citet{bockelee_comet_2017, bockelee_comet_2018} found evidence that during outbursts small (about 100\,nm sized) particles either bound in fractal agglomerates or as individual particles can be ejected. Alice detected enhanced dust densities~\citep{steffl_dust_2015}, and a so called ‘anomalous feature’ suggested to stem from dust particles that are disrupted to fragments in the nanometre size range when entering the instrument~\citep{noonan_investigation_2016, noonan_effects_2016}. Additionally, the langmuir probe of the RPC instrument (Rosetta Plasma Consortium~\citep{carr_rpc_2007}) detected a lack of photoelectrons from the sunward direction over perihelion. One interpretation is the existence of nanometre-sized dust particles between the comet and Sun~\citep{johansson_rosetta_2017}. 
				
				The indirect detections of about 100 nanometre or smaller sized dust particles discussed above suggests that the 100\,nm sized features detected by MIDAS may indeed represent subunits. It might also be hypothesized that the larger, tens to thousands of micrometre-sized dust particles can release their subunits in the nanometre size range under special conditions, e.g., during an outburst, close to strong electric fields, or after a longer Solar System travel.
				
				In conclusion, the sizes of the smallest features detected by MIDAS of about 100\,nm are in good agreement with indirect detections of smallest dust particles by other Rosetta instruments. Thus, these smallest features of MIDAS dust particles might not only be surface related, but may represent subunits.
				
				\paragraph{Interstellar dust grains}
				
				It is an open question to what extent dust particles inherited pristinely from the interstellar medium 
				were available as fundamental building blocks in our early Solar System.
				Based on remote observations, it is expected that interstellar dust consists of silicates and carbon with sizes mostly around a few hundred nanometres, in a distribution reaching down to a few nanometres and up to some micrometres~\citep{li_dust_2003}. 
				
				The cometary material available for investigation typically shows a very small fraction of identifiable interstellar particles. The Stardust collection held a few candidates for interstellar dust~\citep{westphal_interstellar_2014, brownlee_comet_2006}, which were mostly complex aggregates with sizes between some micrometres (for those collected in the aerogel), and a few hundred nanometres (for those captured in the aluminium foils).
				CP IDPs show compositions in agreement with Solar System provenance~\citep{flynn_dust_2016}, but a minority, as small as a few parts per million, of micrometre-sized particles shows isotopic ratios suggesting an interstellar origin~\citep{messenger_material_2000}. Comparing with these low abundances of interstellar particles in cometary dust, it is unlikely that MIDAS particles contain a substantial fraction of interstellar grains. Although the feature sizes identified for particle G would match the expected size range of interstellar particles, due to the lack of compositional data it remains unanswered to which extent MIDAS particles contain interstellar dust.

				
				\subsection{MIDAS dust classification}\label{sec:discussion_classification}
				
				The classification presented in this paper is meant to give a coarse overview of MIDAS results and an easy approach for comparisons to studies about cometary dust and of other Rosetta (dust analysis) instruments~\citep{levasseur_dust_2018}. 
				It is also in agreement with the synthesis of our knowledge about cometary dust that can be found in~\citet{guettler_synthesis_2019}.
				
				\subsubsection{Comparison to Rosetta dust results}
				
				In a tentative combination of different results of the dust analysing instruments on-board Rosetta, namely COSIMA, GIADA, and MIDAS, the majority of the dust of comet 67P in the micro- to millimetre range might be (hierarchical) agglomerates with intermediate porosities around 60 to 90 percent and related densities around 800$\mathrm{\,kg\,m^ {-3}}$~\citep{fulle_dust_2017, langevin_optical_2017}.
				The large compact agglomerates with moderate packing of the subunits at the surface that are mainly detected by MIDAS might be representatives of this group of cometary dust particles detected by COSIMA and GIADA, however, it should be noted that MIDAS large agglomerates are about 10\,$\mathrm{\mu m}$, which is one order of magnitude smaller than the particles usually investigated by COSIMA and GIADA.
				In addition to a majority of compact agglomerates with moderate packing, COSIMA and GIADA find a large dispersion of density values for dust particles of comet 67P: on the one hand, the fluffy fractal particles detected by GIADA and MIDAS show lowest density values less than 1$\mathrm{\,kg\,m^ {-3}}$~\citep{fulle_dust_2017}.
				On the other hand, 
				rather high densities are reached for consolidated, possibly solid particles with densities over 4000$\mathrm{\,kg\,m^ {-3}}$~ detected by GIADA~\citep{fulle_dust_2017}, and by solid and crystalline material detected by COSIMA via their measurement of calcium-aluminium-rich inclusions~\citep{paquette_cai_2016} and due to their detection of crystalline material via specular reflections~\citep{langevin_optical_2017}. 
				However, no solid particles have been detected by MIDAS and the reasons will be discussed in the following section. 
				
				
				\subsubsection{Lack of solid particles}\label{sec:discussion_lack_solid}
				
				As an AFM MIDAS cannot probe the interior of the dust, thus a 'solid' particle for MIDAS is defined as one that does not fragment and that does not show a surface with major features like deep trenches between bulbous units if scanned with sufficient resolution. Typically, large particles show fragmentation and small particles show distinct surface structures. It cannot be excluded that MIDAS scanned particles that had a solid core with a distinct surface layer of bulbous subunits, however, there were no indications for this case. 
				
				A special subgroup of solid particles is represented by euhedral crystalline material. Despite COSIMA’s suggestion of a common admixture of 5 to 15\,$\mathrm{\mu m}$~sized euhedral crystals in their about 100\,$\mathrm{\mu m}$~sized agglomerates~\citep{langevin_optical_2017}, MIDAS did not identify any clear cut crystal shapes. Although a disguise by a surface layer would be conceivable, such a layer would have to be brought in agreement with the detection by COSIMA via specular reflections; one possibility might be a surface layer with pores large enough to allow reflection but small enough to hinder the access with MIDAS’ tips. 
				The same obstacles apply to the detection of euhedral crystals in the smallest subunits found by MIDAS. Although there is no hint for crystalline material at the smallest scale, with the current availability of data a clear decision concerning the amorphous or crystalline nature of the subunits cannot be taken.
				
				Apart from the difficulties in identifying solid particles, their lack in MIDAS' collection could be caused by a lower capture efficiency: solid particles are expected to have a higher probability than agglomerate particles to bounce back rather than stick on the collection target. As the same effect goes for COSIMA collections, the detection of solid particles might be best possible with GIADA. 
				However, 
				GIADA can only detect the density of particles with estimated sizes about 60 to 150\,$\mathrm{\mu m}$ (depending on particle albedo and density~\citep{dellacorte_giada_2015}), sizes slightly above the detection capabilities of MIDAS. Assuming that solid particles are particles with small cross sections and high densities (over 4000$\mathrm{\,kg\,m^ {-3}}$), GIADA found a subset of solid particles that might have no clear counterpart in the particle collections of MIDAS and possibly also not in that of COSIMA~\citep{hilchenbach_mechanical_2017}. It is unknown how many solid particles should be expected at MIDAS' size scale. In conclusion, MIDAS data do not show evidence for solid particles, however, this could be caused by an instrumental bias like collection efficiencies or difficulties with proper identification.

				
				\subsubsection{Fragmentation and hierarchy}\label{sec:hierarchy}
				
				Fragmentation is observed for large (larger than about 10\,$\mathrm{\mu m}$) particles, but no small (less than about 5\,$\mathrm{\mu m}$) ones. It is assumed that the force holding together the grains in the subunits is stronger than the force holding together the subunits of the particle~\citep{hornung_assessment_2016,skorov_dust_2012}. Consequently, fragmentation of the particle in subunits is easier than fragmentation of the subunits in grains. This behaviour is in good agreement with the determined hierarchical agglomerate structure. It is unknown how many levels the hierarchy of cometary dust spans, but MIDAS detected smallest features of about 100\,nm on subunits with sizes of a few hundred nanometres that comprise particles of about a few micrometres size that build agglomerates of about tens of micrometres in size. Additionally, COSIMA inferred the existence of particles up to the millimetre scale~\citep{levasseur_dust_2018, langevin_typology_2016}. Combining these results, dust particles at comet 67P show distinct features at scales between 100\,nm and 1\,mm.
				

				\subsection{Dust of comet 67P: similarities and differences to CP IDPs}\label{sec:comparison}

			\begin{figure*}
				\centering
				\includegraphics[width=12cm]{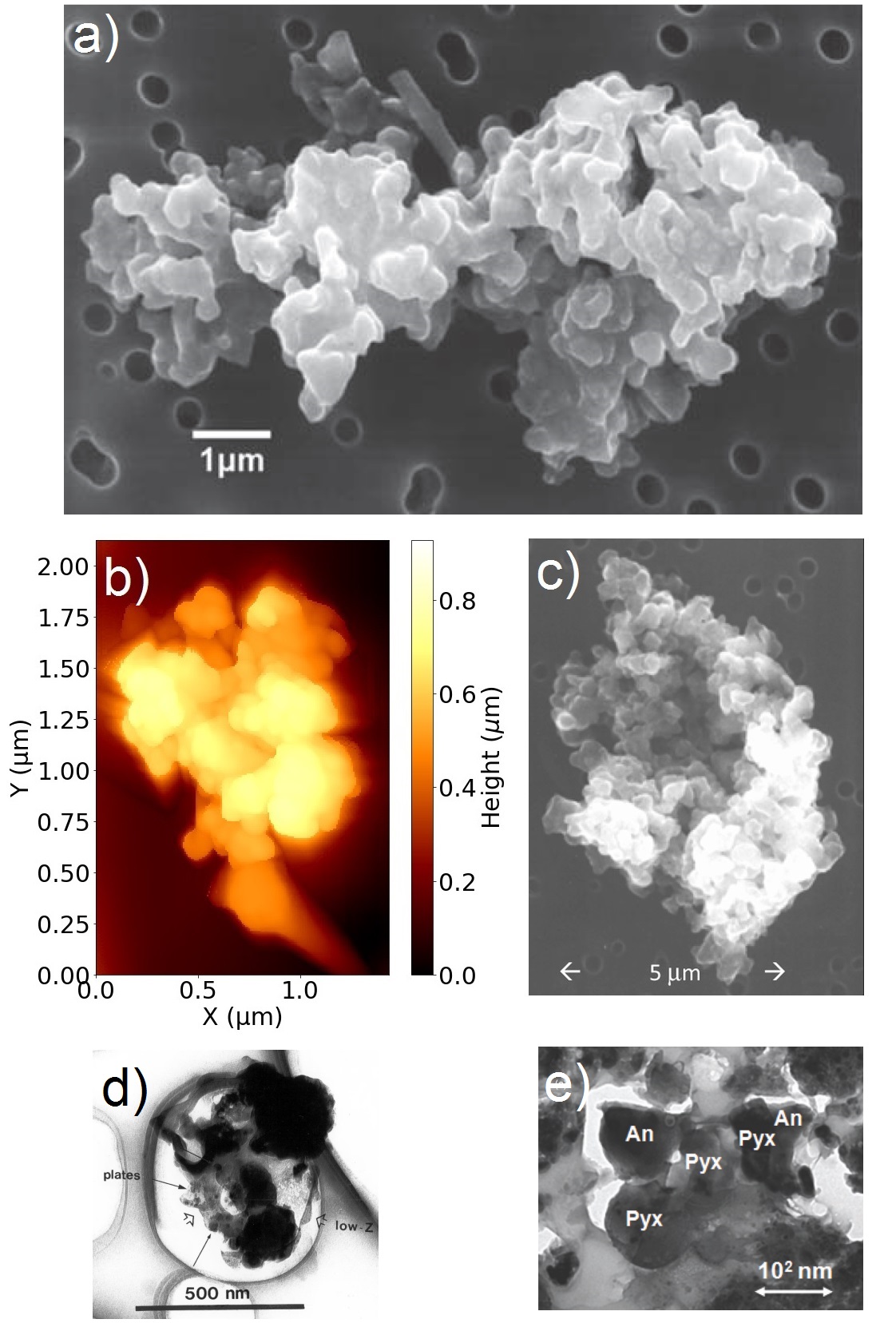}
				\caption{Comparison of the MIDAS particle G in (b) with CP IDPs in (a) and (c), and with high resolution images of their constituent subunits in (d) and (e). (a) is taken from \citet{brownlee_dust_2016}, (c) from \citet{flynn_organic_2013}, (d) from ~\citet{rietmeijer_grain_2004}, and (e) from~\citet{wozniakiewicz_grain_2012}.}
				\label{fig:comp_IDP_grains}
			\end{figure*}

						Although the origin of CP IDPs remains unknown, they have been strongly suggested to stem from comets based on targeted collections during meteor showers linked to comets~\citep{taylor_dust_2016}, modelling of particle trajectories~\citep{poppe_model_2016, nesvorny_origin_2010}, and compositional, optical and structural analysis~\citep{flynn_dust_2016, rietmeijer_idp_1998}.
						Jupiter-family comets were also found to be the main sources of dust particles in Earth's orbit~\citep{levasseur_phase_2019}. 
					
						Results of the Rosetta mission can be used to further strengthen this link. The COSIMA instrument provided compositional and optical analysis of their tens to hundreds of micrometre-sized cometary dust particles. The composition at the 40\,$\mu m$ scale is a mixture of carbonaceous material and minerals, in agreement with CP IDPs and UCAMMs
						~\citep{bardyn_dust_2017}. 
						The appearance of the dust particles collected by MIDAS and COSIMA at comet 67P is also highly reminiscent of CP IDPs~\citep{levasseur_dust_2018, bentley_morphology_2016, langevin_typology_2016}.
						Fig.~\ref{fig:comp_IDP_grains} illustrates these similarities. 
						(a) and (c) show scanning electron microscope images of 8\,$\mu m$ and 11\,$\mu m$ sized CP IDPs. Their surfaces are dominated by bulbous, \mbox{(sub-)}micrometre-sized subunits~\citep{brownlee_dust_2016,noguchi_dust_2015,flynn_organic_2013}. Their surfaces are reminiscent to those of the particles detected by MIDAS, e.g. that of particle G shown in a MIDAS AFM image in (b). However, it should be noted that particle G measures only 1\,$\mu m$, a fraction of the size of typical CP IDPs.
						Fig.~\ref{fig:comp_IDP_grains} (d) and~(e) show transmission electron microscopy images of the subunits in CP IDPs reminiscent to those imaged by MIDAS. (d) presents a mix of silicate features consisting of tiny platy subunits (black arrows) and larger grains (black areas) in a matrix of volatile carbonaceous material (open arrows)~\citep{rietmeijer_grain_2004}. 
						(e) shows polycrystalline silicate grains in a CP IDP 
						with sizes around 100\,nm~\citet{wozniakiewicz_grain_2012}. 
						In a visual comparison, their shapes and assembly are similar to those found in particle G. It should however be noted that no unambiguous comparison can be drawn as no compositional data are available for MIDAS particles. 
					
						In summary, dust of comet 67P and CP IDPs show a similar appearance and their smallest subunits are reminiscent in size and arrangement.
						A promising future project may be a quantitative comparison of the optical images of CP IDPs and cometary dust particles to further strengthen the suggested link between cometary dust and CP IDPs.


					\section{Conclusions}\label{sec:conclusions}
					
					The MIDAS atomic force microscope allowed a morphological classification of nearly pristine cometary dust at the micro- and nanometre scale. 
					Three classes were introduced, namely 
					\begin{itemize}
						\item (i) the large agglomerates of about 10\,$\mathrm{\mu m}$ that consist of micrometre-sized subunits in a fragile arrangement with moderate packing density of the subunits at the surface;
						
						\item (ii) the large porous agglomerate of about 10\,$\mathrm{\mu m}$ comprised by micrometre-sized subunits in a structure with a fractal dimension less than two; 
						
						\item (iii) the small particles of about 1\,$\mathrm{\mu m}$ that show subunits measuring several hundred nanometres with surface features showing a mean size about 100\,nm. 
					\end{itemize}
					
					The MIDAS dust categories are in good agreement with the results found by other dust detecting instruments on-board Rosetta, and in particular the sub-micrometre results allow a good extension of the knowledge of cometary dust to the nanometre scale. 
					
					The nature of the 100\,nm sized surface features of the small particles cannot be conclusively determined by MIDAS data alone, such that the possibility remains that the next larger subunits of some hundreds nanometres size are the fundamental building blocks, or, in contrary, that the fundamental building blocks were not yet detected as they might be even smaller than the smallest features, i.e. less than about 50\,nm in size.  
					
						The size distributions of the smallest detected surface features were determined, where the differential size distribution was found to follow a log-normal distribution with a mean of about 100\,nm and a standard deviation between 20 and 35\,nm. 
						The subunit sizes are in agreement with indirect measurements of other Rosetta instruments. 
						If the subunits found in Stardust material or UCAMMs follow similar size distributions is in the range of possibilities and should be investigated in future projects. 
						CP IDPs show a subunit arrangement, shape and size distribution similar to dust of comet 67P which further strengthens the link between comets and CP IDPs. 
						It also indicates that the smallest, 100\,nm sized features detected by MIDAS might indeed be subunits, however, it remains uncertain if they represent the fundamental building blocks of comet 67P.

	
	\paragraph{\footnotesize{Acknowledgements}}
		 \footnotesize{Rosetta is an ESA mission with contributions from its member states and NASA. We thank the Rosetta Science Ground Segment and Mission Operations Centre for their support in acquiring the data. 
		 MIDAS became possible through support from funding agencies including the European Space Agency’s PRODEX programme, the Austrian Space Agency, the Austrian Academy of Sciences and the German funding agency DARA	 (later DLR).
		 T. Mannel acknowledges funding by the Austrian Science Fund FWF P 28100-N36 and A.C. Levasseur-Regourd acknowledges support from Centre National d’Études Spatiales in the scientific analysis of the Rosetta mission.}

\bibliographystyle{apalike}
\bibliography{AA_MIDAS_bib}

\renewcommand{\thesection}{\Roman{section}} 
\renewcommand{\thesubsection}{\thesection.\Roman{subsection}}
\setcounter{section}{0}

\section*{Appendix}
\section{Dust collection time, exposure geometry, and scan history related to particle G}\label{history}

Unlike regular MIDAS scans where a clear strategy of scan, expose and re-scan was usually employed, the determination of when a tip picked-up dust, from what target it originated, and when this dust was first collected requires a careful analysis. Calibration images of the tip in use were commanded periodically (typically once every few weeks) to monitor the tip health and, towards the end of the mission, specifically for the purpose of high resolution reverse imaging as described in this paper. After identifying a dust particle sticking to a tip, all exposures and operations in between have to be examined to see which was the likely period of dust collection and the time of the pick-up.

\begin{figure*}
	\centering
	\includegraphics[width = 17cm]{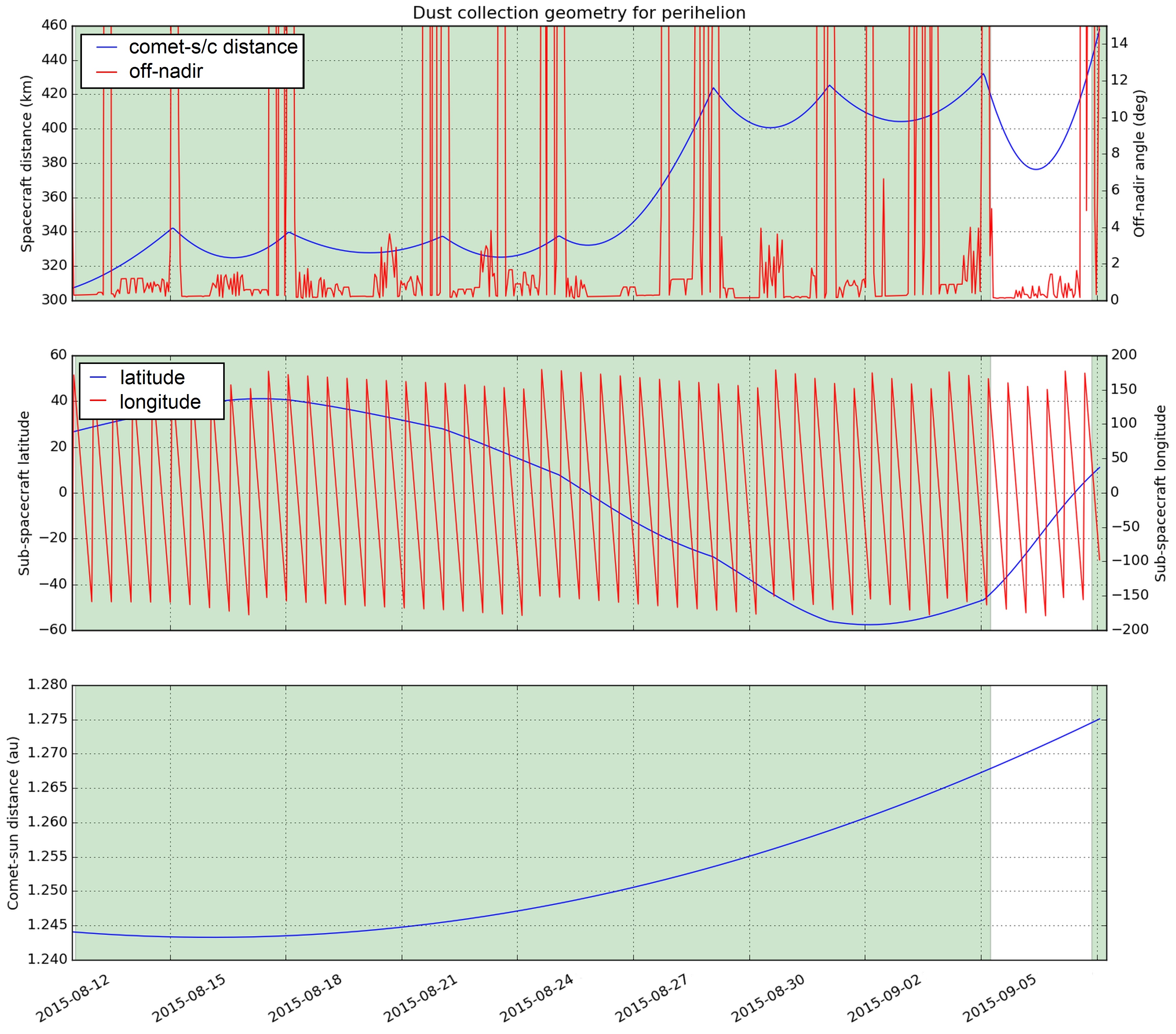}
	\caption{The collection geometry for particle G (see Figs.~\ref{fig_small_particle_15} and~\ref{fig_small_particle_8}). Green shaded regions show the periods when particles could be collected. The upper panel gives the comet-spacecraft distance and off-nadir pointing, the middle panel shows latitude and longitude, and the bottom panel the comet-sun distance.}
	\label{fig:collection_geom}
\end{figure*}

The tip shown in Fig.~\ref{fig_small_particle_15} and~\ref{fig_small_particle_8} was only used for scans on a target exposed during perihelion. The exact exposure geometry is shown in Fig.~\ref{fig:collection_geom}. It can be seen that the spacecraft had a distance to the comet between 300 and 450\,km while the comet was at its closest position to the Sun at about 1.2\,au. Following the orientation of the Rosetta probe, MIDAS funnel was pointing at various possible locations around the whole comet with latitudes between +40 and -60 degrees and longitudes between $\pm$ 50 degrees. 

The successful images taken with the tip in the normal imaging mode have resolutions up to 625\,nm and do not show obvious dust particles. However, it is still possible that small dust particles that were not visible in these coarse resolution scans were existent and (partly) picked-up. 
The tip image was taken after only four scans of this target and shows a large dust particle attached to the tip. Unfortunately, no previous	tip image is available for this tip and thus the cometary origin of the contaminant must be concluded based on the morphological similarities to other dust particles adhering to tips and detected on the targets.

\section{Discussion of uncertainties}\label{sec:uncertainties}

Although it is tempting to interpret AFM images as one would their optical counterparts, care must be taken as every AFM image is a convolution of the tip and sample shapes. 
Since the tip convolution is strongly dependent on the tip opening angle, the artefacts are reduced by using the reverse imaging technique with the calibration standard spikes. In addition, different areas on the calibration sample (with different spikes) were used, and typically the tip was imaged by more than one spike within a given scan. This gave confidence that the features seen in these images were not the result of contamination on the tip calibration sample. 

To calculate uncertainties for the feature sizes two sources of inaccuracies were taken into account. 
First, the tip-sample-convolution leads to a systematic broadening of the measured size. Second, the marking precision of the features introduces a statistical deviation of the size. Systematical and statistical errors are added linearly to arrive at the total uncertainty. As the tip convolution can only broaden the size of positive features, but uncertainties in the marking of the subunits can either enlarge or shrink the sizes, the resulting error bars are asymmetric.

The degree of tip convolution depends on the curvature radius of the calibration spike, its opening angle, and the shape of the feature to be imaged. The increase in the diameter $\Delta d$ is approximated as $\Delta d = tan\left( \frac{\alpha}{2} \right) \cdot (h_{mean} - h_{min})$, where $\frac{\alpha}{2}$ is the half opening angle of the tip of 25$^{\circ} \pm $ 5$^{\circ}$, and the mean and minimal height measured under the marked area are $h_{mean}$ and $h_{min}$. A small height difference $h_{mean} - h_{min}$ leads to a small tip convolution, where the minimal reachable value is given by the apex diameter of the spikes on the calibration target of 20\,nm~\citep{cal_sample}. The presented calculation is precise for spherical features and will result in an overestimation for more pointed or flatter objects. It thus gives an upper limit of the uncertainty. The uncertainty calculated ranges between 14 and 234 percent,
	where the smallest values originate from the flattest subunits. Their flanks are shallower than the tip half-opening angle and thus the broadening is limited to the size of the tip apex (for a 100\,nm sized subunit this would be 20~percent uncertainty due to tip convolution). 
The extreme values (over about 80 percent) occur for features located at the rims of the particles. As the sides of the dust sticking to a tip are not in contact with a target surface but instead adhere to the wall of the tip, the spike on the calibration target can approach from the side and thus create the impression of an arbitrarily lengthened feature. 
The resulting large height differences lead to correspondingly large tip convolution uncertainties. Those cases are clearly identifiable in the cumulative size distributions due to their large  uncertainty bars; those error bars that would otherwise have extended, non-physically, below zero diameter have been truncated at zero. 

The marking precision of features in the scans depends on the resolution of the scan and the accuracy of the recognition if a pixel belongs to a feature or not (i.e. a human factor). To estimate the latter
uncertainty, the features were repeatedly marked and the results compared. This lead to an assumption of a 15~percent deviation in the number of marked pixels. 
	Calculating the deviation of the measured diameter on this basis leads to uncertainties between 7 and 9 percent. 
Again, especially those features close to a rim produce the higher error rates. However, as evenly distributed erroneously marked pixels do not have a strong influence on the calculated feature size, the uncertainty due to the marking is in the range of the image resolution or smaller. 

	Some relatively large uncertainties for the subunit sizes present a challenge when determining the differential size distribution, as the binning cannot be chosen smaller than the maximal uncertainty found for one of the data points. 
	For the presented cases the differential size distribution would only contain 2 bins, which renders the determination of a distribution function impossible.  
	However, the cumulative size distribution does not face these problems as here no binning is necessary. Thus, this paper will investigate the shape of the cumulative size distribution and infer properties of the differential size distribution.

\section{Fit of the size distributions}\label{sec_app_fit}

	The cumulative size distribution of the subunits is expected to follow the integrand of the log-normal distribution
	\begin{equation}
	\begin{split}
	a\cdot &\int \frac{1}{\sqrt{2\pi} s x} exp\bigg(- \frac{(ln(x) - m) ^2}{2s ^2}\bigg) dx \\
	&=  \frac{a}{2} \cdot\bigg(1 + erf\bigg( \frac{ln(x) - m}{\sqrt{2}s} \bigg)\bigg).
	\end{split}
	\label{eq_lognormal}
	\end{equation} 
	The related mean value $\mu_{log}$ and standard deviation $\sigma_{log}$ for the log-normal distribution are calculated as
	\begin{equation}
	\mu_{log} = e^{m+\frac{s^2}{2}},
	\label{eq_transf_m}
	\end{equation}
	\begin{equation}
	\sigma_{log} = e^{m + \frac{s^2}{2}}\cdot \sqrt{(e^{s^2}-1)}.
	\label{eq_transf_s}
	\end{equation}

	All fits were carried out by the orthogonal distance regression routine of python (scipy.odr, \url{https://docs.scipy.org/doc/scipy/reference/odr.html}). 
	The fits take into account the uncertainties of the data and return fitted values together with uncertainties. Uncertainties of derived quantities (the mean and standard deviation of the log-normal distribution, see Eqs.~\ref{eq_transf_m} and~\ref{eq_transf_s}) are propagated, where their contributions are added quadratically.

	The fits were tested by a KS test, where the distance in y-direction between the cumulative size distributions and their empirical distribution functions were calculated. 
	In all cases the determined distances where well below the maximally allowed distance to pass the test (for the 15\,nm log-normal fit $d_{15nm} = 0.057$ < $d_{max\_15nm} = 0.116$, for the 8\,nm log-normal fit $d_{8nm} = 0.045$ < $d_{max\_8nm} = 0.117$, for the line fit of particle G $d_{G} = 0.1$ < $d_{max\_G} = 0.3$, and for particle D $d_{D} = 0.1$ < $d_{max\_D} = 0.4$).

\section{Open access of MIDAS data, raw images, and image processing}\label{sec:raw}

The data used in this paper are available in the ESA Planetary Science Archive at \url{https://archives.esac.esa.int/psa/\#!Table\%20View/MIDAS=instrument} with the product identifiers as given in Tab.~\ref{table:meta_15} and~\ref{table:meta_8}. The tables also contain the key metadata of the scans.

Figs.~\ref{fig_small_particle_15} and~\ref{fig_small_particle_8} only contain a crop of the data, thus the full scans are shown in Figs.~\ref{fig:raw_15} and~\ref{fig:raw_8}.
It is obvious that the images show a slight wavelike bending in x-direction, an effect due to thermal drift of the piezo motor (see, e.g.~\citep{eaton_atomic_2010}). 
MIDAS was originally foreseen to operate in a closed loop design to remove artefacts due to the behaviour of the piezo motors including the above mentioned thermal drift, however, this function was lost during launch for the x-direction~\citep{bentley_lessons_2016}. 
The scans presented in this paper are taken with y as fast scanning direction (from top to bottom) in a closed loop to allow correction for piezo creep, and x as slow scanning direction (from left to right) in an open loop. Thus, the temperature drift is especially strong in x direction and for longer scanning durations.  

In principle, it is possible to remove the wavelike bending by a polynomial background subtraction. This procedure adjusts the height of the data, but not the stepsize in the x- and y-directions. This paper analyses the sizes of the particle and its subunits projected on the x-y-plane and does not use the height information other than for feature identification. Since every processing step alters the data and can introduce a bias, it was decided to apply no such processing. 	 
To correct deviations of the stepsize in the x- and y-directions, a dedicated calibration scan would have been necessary, but due to the complex planning pattern of MIDAS~\citep{bentley_lessons_2016} this was not feasible for the herein presented scans. The equivalent diameters of particle G in the analysed scans are with $1213^{+32}_{-390}$\,nm and $1255^{+37}_{-460}$\,nm for the 15\,nm and 8\,nm scan, respectively, similar in the range of uncertainties despite the slight bending. As the error introduced by the thermal drift is even less for the smaller subunits, it is assumed that the unprocessed data are still a valid basis for the presented data analysis.

\begin{minipage}{\linewidth}
\begingroup
	\centering
	\includegraphics[width = \linewidth]{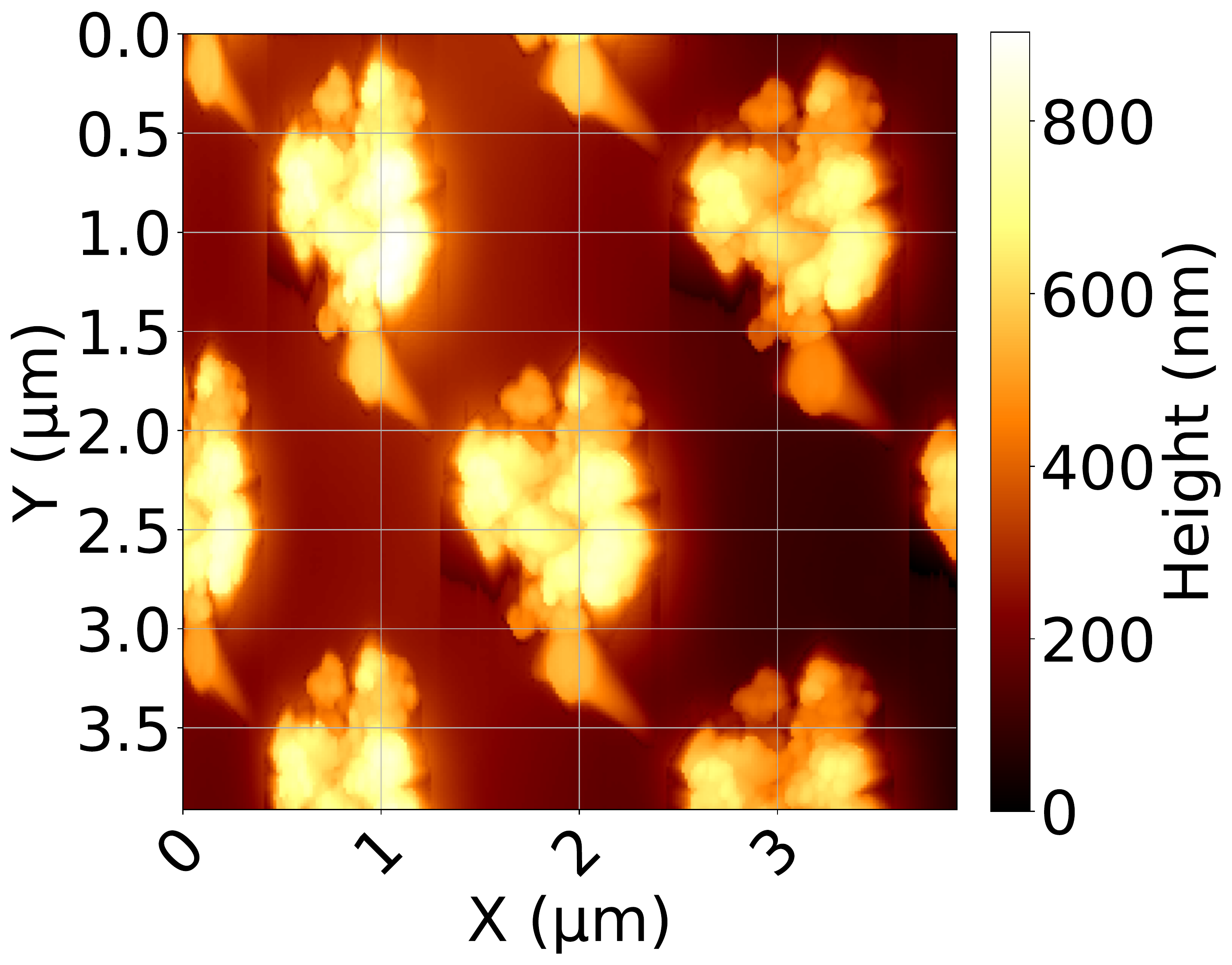}
	\captionof{figure}{Full image of the scan taken on 08 December 2015 shown in Fig.~\ref{fig_small_particle_15}. The key metadata are given in Tab.~\ref{table:meta_15}.}
	\label{fig:raw_15}
\endgroup
\end{minipage}

\begingroup
\begin{minipage}{\linewidth}
	\centering
	\includegraphics[width = \linewidth]{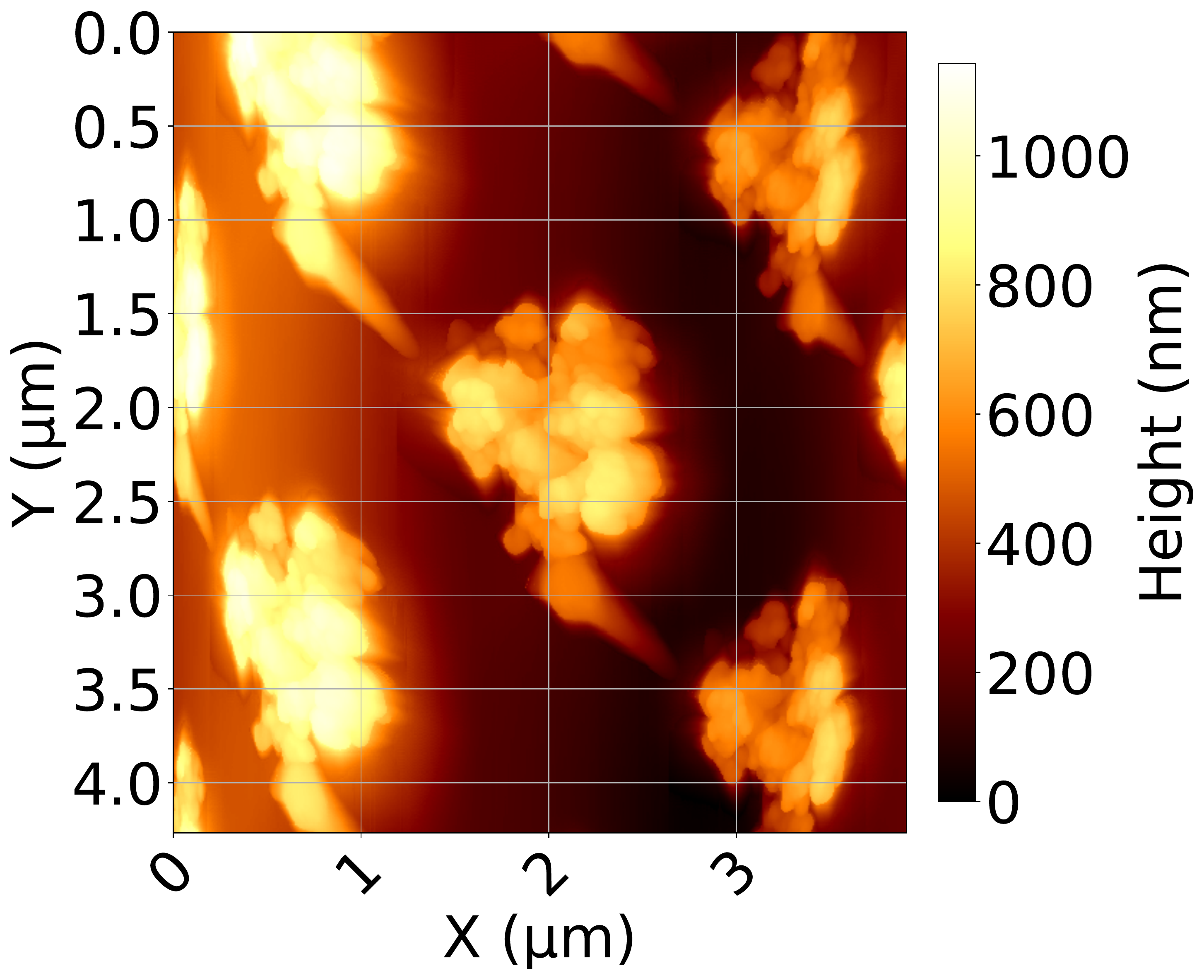}
	\captionof{figure}{Full image of the scan taken on 11 May 2016 shown in Fig.~\ref{fig_small_particle_8}. The key metadata are given in Tab.~\ref{table:meta_8}.}
	\label{fig:raw_8}
\end{minipage}
\endgroup

	\vspace{6cm}
	\begingroup
	\begin{minipage}{\linewidth}
		\captionof{table}{Metadata of the scan shown in Figs.~\ref{fig_small_particle_15} and~\ref{fig:raw_15}.} 
		\label{table:meta_15} 
		\centering 
		\begin{tabular}{l l} 
			\hline 
			archive dataset & RO-C-MIDAS-3-ESC4-SAMPLES-V2.0 \\ 
			archive product ID & IMG\_1532123\_1535000\_076\_ZS \\
			scan start time & 2015-12-08 12:34:27 UTC\\
			duration & 7:16:18 \\
			x resolution & 15.3\,nm \\
			y resolution & 15.3\,nm \\
			z resolution & 0.7\,nm \\
			fast scan direction & y (top to bottom) \\
			slow scan direction & x (left to right) \\
			tip number & 15 \\
			target number & 04 (tip calibration)\\
			\hline 
		\end{tabular}
		\end{minipage}
	\endgroup
	
	\vspace{3cm}
	
	\begin{minipage}{\linewidth}
	\begingroup
		\captionof{table}{Metadata of the scan shown in Figs.~\ref{fig_small_particle_8} and~\ref{fig:raw_8}.} 
		\label{table:meta_8} 
		\centering 
		\begin{tabular}{l l} 
			\hline 
			archive dataset & RO-C-MIDAS-3-EXT2-SAMPLES-V2.0 \\ 
			archive product ID & IMG\_1612423\_1615300\_043\_ZS \\
			scan start time & 2016-05-11 12:09:28 UTC \\
			duration & 22:22:57 \\
			x resolution & 7.6\,nm \\
			y resolution & 8.3\,nm \\
			z resolution & 0.7\,nm \\
			fast scanning direction & y (top to bottom) \\
			slow scanning direction & x (left to right) \\
			tip number & 15 \\
			target number & 04 (tip calibration)\\
			\hline 
		\end{tabular}
	\endgroup
	\end{minipage}

\end{multicols}	

\newpage
\renewcommand{\thesection}{\Roman{section}} 
	\section{Tabulated subunit sizes}
	
	The sizes of the subunits and features identified in Fig.~\ref{fig_small_particle_15} (b) and (c), and in Fig.~\ref{fig_small_particle_8} (b) are given with their uncertainties in Tab.~\ref{15_large}, Tab.~\ref{15_small}, and Tab.~\ref{8_small}, respectively.

\begin{longtable}{ccc}
	\caption{\label{15_large} Tabulated sizes of the subunits of particle G as shown in Fig.~\ref{fig_small_particle_15} (b) and in Fig.~\ref{fig:cum_sd_large_15}.}\\
	\hline\hline
	d (nm) & +$\Delta$d (nm) (+$\Delta$d (\%)) & -$\Delta$d (nm)  (-$\Delta$d (\%)) \\
	\hline
	\endfirsthead
	\caption{continued.}\\
	\hline\hline
	d (nm) & +$\Delta$d (nm) (+$\Delta$d (\%)) & -$\Delta$d (nm)  (-$\Delta$d (\%)) \\
	\hline
	\endhead
	\hline
	\endfoot
	\endlastfoot
	271 & 21 (8 \%) & 155 (57 \%) \\
	280 & 21 (8 \%) & 185 (66 \%) \\ 
	350 & 26 (8 \%) & 233 (67 \%) \\ 
	350 & 26 (8 \%) & 233 (67 \%) \\ 
	450 & 34 (8 \%) & 303 (67 \%) \\ 
	456 & 35 (8 \%) & 242 (53 \%) \\ 
	482 & 36 (8 \%) & 314 (65 \%) \\ 
	555 & 42 (8 \%) & 355 (64 \%) \\ 
\end{longtable}

\begin{longtable}{ccc}
	\caption{\label{15_small} Tabulated sizes of the subunits of particle G as shown in Fig.~\ref{fig_small_particle_15} (c) and in Fig.~\ref{fig:cum_sd_small_15}.}\\
	\hline\hline
	d (nm) & +$\Delta$d (nm) (+$\Delta$d (\%)) & -$\Delta$d (nm)  (-$\Delta$d (\%)) \\
	\hline
	\endfirsthead
	\caption{continued.}\\
	\hline\hline
	d (nm) & +$\Delta$d (nm) (+$\Delta$d (\%)) & -$\Delta$d (nm)  (-$\Delta$d (\%)) \\
	\hline
	\endhead
	\hline
	\endfoot
	\endlastfoot
	52 & 6 (11 \%) & 26 (50 \%) \\ 
	52 & 6 (11 \%) & 26 (50 \%) \\ 
	55 & 5 (10 \%) & 25 (47 \%) \\ 
	57 & 5 (9 \%) & 139 (243 \%) \\ 
	60 & 5 (8 \%) & 30 (51 \%) \\ 
	65 & 7 (11 \%) & 27 (42 \%) \\ 
	65 & 7 (11 \%) & 27 (42 \%) \\ 
	71 & 6 (9 \%) & 26 (37 \%) \\ 
	73 & 6 (8 \%) & 44 (60 \%) \\ 
	75 & 6 (8 \%) & 26 (35 \%) \\ 
	77 & 6 (8 \%) & 40 (51 \%) \\ 
	79 & 8 (10 \%) & 28 (35 \%) \\ 
	79 & 8 (10 \%) & 32 (41 \%) \\ 
	79 & 8 (10 \%) & 28 (35 \%) \\ 
	83 & 7 (9 \%) & 27 (33 \%) \\ 
	84 & 7 (8 \%) & 27 (32 \%) \\ 
	86 & 7 (8 \%) & 55 (64 \%) \\ 
	88 & 7 (8 \%) & 30 (34 \%) \\ 
	88 & 7 (8 \%) & 85 (97 \%) \\ 
	88 & 7 (8 \%) & 96 (109 \%) \\ 
	88 & 7 (8 \%) & 51 (58 \%) \\ 
	88 & 7 (8 \%) & 27 (31 \%) \\ 
	90 & 8 (9 \%) & 41 (45 \%) \\ 
	90 & 8 (9 \%) & 49 (55 \%) \\ 
	91 & 8 (9 \%) & 33 (37 \%) \\ 
	91 & 8 (9 \%) & 72 (80 \%) \\ 
	93 & 8 (9 \%) & 28 (30 \%) \\ 
	93 & 8 (9 \%) & 36 (39 \%) \\ 
	93 & 8 (9 \%) & 32 (34 \%) \\ 
	93 & 8 (9 \%) & 47 (50 \%) \\ 
	98 & 8 (8 \%) & 51 (53 \%) \\ 
	98 & 8 (8 \%) & 36 (37 \%) \\ 
	99 & 8 (8 \%) & 48 (49 \%) \\ 
	101 & 9 (9 \%) & 33 (33 \%) \\ 
	101 & 9 (9 \%) & 61 (60 \%) \\ 
	101 & 9 (9 \%) & 29 (29 \%) \\ 
	101 & 9 (9 \%) & 92 (92 \%) \\ 
	103 & 9 (8 \%) & 50 (48 \%) \\ 
	103 & 9 (8 \%) & 29 (28 \%) \\ 
	103 & 9 (8 \%) & 51 (49 \%) \\ 
	103 & 9 (8 \%) & 29 (28 \%) \\ 
	103 & 9 (8 \%) & 55 (54 \%) \\ 
	105 & 9 (8 \%) & 53 (50 \%) \\ 
	108 & 8 (8 \%) & 41 (38 \%) \\ 
	108 & 8 (8 \%) & 85 (79 \%) \\ 
	109 & 8 (8 \%) & 66 (60 \%) \\ 
	109 & 8 (8 \%) & 90 (83 \%) \\ 
	110 & 9 (9 \%) & 29 (27 \%) \\ 
	110 & 9 (9 \%) & 64 (58 \%) \\ 
	114 & 9 (8 \%) & 48 (42 \%) \\ 
	116 & 9 (8 \%) & 93 (81 \%) \\ 
	116 & 9 (8 \%) & 42 (37 \%) \\ 
	116 & 9 (8 \%) & 46 (40 \%) \\ 
	118 & 10 (9 \%) & 35 (29 \%) \\ 
	118 & 10 (9 \%) & 105 (89 \%) \\ 
	119 & 10 (8 \%) & 55 (46 \%) \\ 
	123 & 10 (8 \%) & 42 (34 \%) \\ 
	125 & 10 (8 \%) & 119 (95 \%) \\ 
	125 & 10 (8 \%) & 52 (41 \%) \\ 
	125 & 10 (8 \%) & 45 (36 \%) \\ 
	127 & 11 (8 \%) & 69 (55 \%) \\ 
	130 & 10 (8 \%) & 74 (57 \%) \\ 
	130 & 10 (8 \%) & 50 (39 \%) \\ 
	138 & 11 (8 \%) & 62 (45 \%) \\ 
	144 & 11 (8 \%) & 37 (26 \%) \\ 
	152 & 12 (8 \%) & 87 (57 \%) \\ 
	165 & 13 (8 \%) & 102 (62 \%) \\ 
	183 & 14 (8 \%) & 139 (76 \%) \\ 
\end{longtable}

\begin{longtable}{c c c}
	\caption{\label{8_small} Tabulated sizes of the subunits of particle G as shown in Fig.~\ref{fig_small_particle_8} (b) and in Fig.~\ref{fig:cum_sd_small_8}.}\\
	\hline\hline
	d (nm) & +$\Delta$d (nm) (+$\Delta$d (\%)) & -$\Delta$d (nm)  (-$\Delta$d (\%)) \\
	\hline
	\endfirsthead
	\caption{continued.}\\
	\hline\hline
	d (nm) & +$\Delta$d (nm) (+$\Delta$d (\%)) & -$\Delta$d (nm)  (-$\Delta$d (\%)) \\
	\hline
	\endhead
	\hline
	\endfoot
	\endlastfoot
	45 & 3 (8 \%) & 23 (52 \%) \\ 
	48 & 4 (9 \%) & 24 (51 \%) \\ 
	49 & 4 (8 \%) & 24 (49 \%) \\ 
	53 & 4 (8 \%) & 24 (47 \%) \\ 
	53 & 4 (8 \%) & 33 (61 \%) \\ 
	56 & 4 (7 \%) & 24 (43 \%) \\ 
	57 & 4 (7 \%) & 28 (49 \%) \\ 
	61 & 4 (7 \%) & 26 (42 \%) \\ 
	64 & 5 (8 \%) & 25 (39 \%) \\ 
	65 & 5 (7 \%) & 25 (38 \%) \\ 
	66 & 5 (7 \%) & 25 (38 \%) \\ 
	68 & 5 (8 \%) & 25 (37 \%) \\ 
	69 & 5 (7 \%) & 25 (37 \%) \\ 
	69 & 5 (7 \%) & 31 (45 \%) \\ 
	71 & 6 (8 \%) & 26 (36 \%) \\ 
	74 & 6 (8 \%) & 33 (45 \%) \\ 
	75 & 6 (8 \%) & 141 (188 \%) \\ 
	75 & 6 (8 \%) & 26 (34 \%) \\ 
	76 & 6 (7 \%) & 26 (34 \%) \\ 
	78 & 6 (8 \%) & 26 (33 \%) \\ 
	79 & 6 (8 \%) & 26 (33 \%) \\ 
	80 & 6 (7 \%) & 38 (48 \%) \\ 
	81 & 6 (7 \%) & 28 (35 \%) \\ 
	82 & 6 (8 \%) & 26 (32 \%) \\ 
	83 & 6 (7 \%) & 29 (34 \%) \\ 
	86 & 6 (7 \%) & 29 (34 \%) \\ 
	88 & 7 (8 \%) & 62 (70 \%) \\ 
	89 & 7 (7 \%) & 27 (30 \%) \\ 
	91 & 7 (7 \%) & 125 (137 \%) \\ 
	92 & 7 (7 \%) & 44 (48 \%) \\ 
	93 & 7 (7 \%) & 27 (29 \%) \\ 
	94 & 7 (8 \%) & 33 (35 \%) \\ 
	94 & 7 (8 \%) & 50 (54 \%) \\ 
	97 & 7 (7 \%) & 77 (79 \%) \\ 
	97 & 7 (7 \%) & 52 (53 \%) \\ 
	98 & 7 (7 \%) & 40 (41 \%) \\ 
	99 & 7 (7 \%) & 41 (41 \%) \\ 
	100 & 7 (7 \%) & 47 (47 \%) \\ 
	102 & 8 (7 \%) & 43 (42 \%) \\ 
	102 & 8 (7 \%) & 36 (36 \%) \\ 
	103 & 8 (7 \%) & 135 (131 \%) \\ 
	103 & 8 (7 \%) & 28 (27 \%) \\ 
	106 & 8 (7 \%) & 45 (43 \%) \\ 
	106 & 8 (7 \%) & 49 (46 \%) \\ 
	108 & 8 (7 \%) & 67 (62 \%) \\ 
	109 & 8 (7 \%) & 142 (131 \%) \\ 
	110 & 8 (7 \%) & 28 (26 \%) \\ 
	116 & 8 (7 \%) & 30 (26 \%) \\ 
	116 & 8 (7 \%) & 59 (51 \%) \\ 
	117 & 9 (7 \%) & 119 (102 \%) \\ 
	118 & 9 (7 \%) & 29 (24 \%) \\ 
	118 & 9 (7 \%) & 36 (31 \%) \\ 
	119 & 9 (7 \%) & 44 (37 \%) \\ 
	120 & 9 (7 \%) & 46 (38 \%) \\ 
	121 & 9 (7 \%) & 49 (40 \%) \\ 
	124 & 9 (7 \%) & 135 (109 \%) \\ 
	127 & 9 (7 \%) & 97 (76 \%) \\ 
	127 & 9 (7 \%) & 126 (99 \%) \\ 
	132 & 10 (7 \%) & 87 (66 \%) \\ 
	143 & 11 (7 \%) & 56 (39 \%) \\ 
	147 & 11 (7 \%) & 135 (92 \%) \\ 
	150 & 11 (7 \%) & 68 (46 \%) \\ 
	154 & 11 (7 \%) & 50 (32 \%) \\ 
	157 & 11 (7 \%) & 33 (21 \%) \\ 
	188 & 14 (7 \%) & 106 (57 \%) \\ 
	212 & 15 (7 \%) & 128 (60 \%) \\ 
	216 & 16 (7 \%) & 157 (72 \%) \\ 
\end{longtable}

\end{document}